\documentclass[a4paper,onecolumn,11pt,accepted=2025-05-28]{quantumarticle} 
\pdfoutput=1

\usepackage[utf8]{inputenc}
\usepackage[english]{babel}
\usepackage[T1]{fontenc}
\usepackage{amsmath}
\usepackage{hyperref}
\usepackage{cite}

\usepackage{bbold} 
\usepackage{graphicx}
\usepackage{bm}
\usepackage{comment}

\newcommand{\beq}{\begin{equation}}
\newcommand{\eeq}{\end{equation}}

\usepackage{color}

\usepackage[normalem]{ulem}
\newcommand\reduline{\bgroup\markoverwith
{\textcolor{red}{\rule[0.5ex]{2pt}{0.4pt}}}\ULon}

\definecolor{mygreen}{rgb}{0.0,0.55,0.3}

\definecolor{calumblue}{RGB}{10, 62, 169}

\definecolor{calumpurple}{RGB}{100, 20, 230}

\definecolor{calumred}{RGB}{180,24,24}

\newcommand{\He}{H_\text{eff}}

\newcommand{\WF}{\mathcal{W}_F}
\newcommand{\WC}{\mathcal{W}_C}
\newcommand{\SJED}{S_\alpha}
\newcommand{\cgo}{\mathcal{A}}

\newcommand{\cset}{\mathbf{C}}

\newcommand{\ui}{{j}} 
\newcommand{\deph}{J}
\newcommand{\pic}{\pi_\textsc{c}}
\newcommand{\pa}{{\pic(\alpha)}}

\newcommand{\Tr}{\text{Tr}}

\newcommand{\dc}{d_C}
\newcommand{\dct}{\tilde{d}_C}

\newcommand{\rs}{\chi}

\makeatletter
\newcommand{\phantomlabel}[2]{
    \protected@write\@auxout{}{
        \string\newlabel{#2}{
            {\@currentlabel#1}{\thepage}
            {\@currentlabel#1}{#2}{}
        }
    }
    \hypertarget{#2}{}
}
\makeatother

\usepackage{hyperref}
\hypersetup{
    colorlinks=true,
    linkcolor=black,
    citecolor=blue,
}

\begin{document}

\title{Gauge freedoms in unravelled quantum dynamics: When do different continuous measurements yield identical quantum trajectories?}

\author{Calum A. Brown}
\email{cb2177@cam.ac.uk}
\affiliation{Department of Applied Mathematics and Theoretical Physics, University of Cambridge, Wilberforce Road, Cambridge\\
CB3 0WA, United Kingdom}
\orcid{0009-0000-8156-7305}
\author{Katarzyna Macieszczak}
\affiliation{Department of Physics, University of Warwick, Coventry CV4 7AL, United Kingdom}
\orcid{0000-0002-9814-164X}
\author{Robert L. Jack}
\affiliation{Department of Applied Mathematics and Theoretical Physics, University of Cambridge, Wilberforce Road, Cambridge\\
CB3 0WA, United Kingdom}
\affiliation{Yusuf Hamied Department of Chemistry, University of Cambridge, Lensfield Road, Cambridge CB2 1EW, United Kingdom}
\orcid{0000-0003-0086-4573}

\maketitle

\hypersetup{
    colorlinks=true,
    linkcolor=magenta,
    citecolor=blue,
}

\begin{abstract}
Quantum trajectories of a Markovian open quantum system arise from the back-action of measurements performed in the environment with which the system interacts. In this work, we consider counting measurements of quantum jumps, and the associated representations of the quantum master equation. 
We derive necessary and sufficient conditions under which different measurements give rise to the same unravelled quantum master equation, which governs the dynamics of the probability distribution over pure conditional states of the system. Since that equation uniquely determines the stochastic dynamics of a conditional state, we also obtain necessary and sufficient conditions under which different measurements result in identical quantum trajectories.  
We then consider the joint stochastic dynamics for the conditional state and the measurement record.  We formulate this in terms of labelled quantum trajectories, and derive necessary and sufficient conditions under which different representations lead to equivalent labelled quantum trajectories, up to permutations of labels. As those conditions are generally stricter, we finish by constructing coarse-grained measurement records, such that equivalence of the corresponding partially-labelled trajectories is guaranteed by equivalence of the trajectories alone.  These general results are illustrated by two examples that demonstrate permutation of labels, and equivalence of different quantum trajectories.

\end{abstract}

\maketitle


\hypersetup{
    colorlinks=true,
    linkcolor=magenta,
    citecolor=blue,
}

\section{Introduction}
\label{sec:intro}

\noindent\emph{Motivation} -- Open quantum systems are important in many physical contexts where the influence of external environments cannot be neglected \cite{Breuer_and_Petruccione,Wiseman2010,Gardiner2004,Fazio_notes}. They are often analysed using the \emph{quantum master equation} (QME) \cite{Lindblad1976,Gorini1976} where the system state is described via a density matrix.  This averaged state follows a  non-unitary evolution, due to the interaction with the environment. In contrast to open classical systems, the QME does not uniquely prescribe stochastic dynamics of fluctuating trajectories of the system.  In fact, a single QME permits many different unravellings, each corresponding to a specific stochastic process for a conditional system's state.  Such processes arises from the back-action of a chosen continuous measurement on the environment.  These possibilities appear in the QME via a set of gauge freedoms: the different stochastic processes are associated to different \emph{representations} of the quantum master operator, corresponding to different decompositions in terms of jump operators and a system Hamiltonian.

The dynamics of \emph{quantum trajectories} has gained increased interest in recent years, covering subjects such as measurement induced (and other) phase transitions \cite{Chan2019,Skinner2019,Szyniszweski2019,Turkeshi2021,Turkeshi2021,Turkeshi2022,LeGal24,Garrahan10,Lesanovsky13,gebhart2020,Biella2021,cabot23}, quantum control \cite{abdelhafez19,Propson21,grigoletto22,herasymenko23}, quantum stochastic thermodynamics \cite{Hekking2013,Manzano2022,Almeida2024} and steady state ensemble preparation \cite{Choi2023,Cotler2023,Ippoliti2023}. Experimentally, quantum trajectories are obtained from continuous monitoring of the system, e.g., by detecting its output into the environment. Advances in experimental techniques have now produced practical platforms in which to investigate phenomena related to these stochastic trajectories \cite{Gleyzes2007,Guerlin2007,Deleglise2008,Murch2013,Hofmann2016,Fink2018,Kurzmann2019,minev19}.
In each case, the representation, which is determined by the continuous monitoring procedure used, plays a crucial role and different choices can have drastic impacts on the dynamics, 
including phenomena such as transport and entanglement entropy, as well as classical simulability \cite{Plenio1998,vovk22,gneiting22,chen24,vovk24,Eissler25}.
Indeed, trajectories play an important role in the numerical simulation of quantum master equations~\cite{Dum1992,Dalibard1992,Molmer93,Plenio1998,Daley2014,Radelli2024,Macieszczak2021,Elliot2024}, which exploit the reduced dimensionality of pure conditional states, as opposed to (generically) mixed density matrices.

With the focus of applications on quantum trajectories and their properties, important questions remain about correspondence between their dynamics and continuous measurements or unravellings.   In particular, while the choice of continuous measurement uniquely defines the quantum trajectory dynamics, the converse does not hold.  This non-uniqueness is the very freedom in the choice of experimental and numerical protocols for the generation of quantum trajectories.

\noindent\emph{Theoretical framework} -- Here, we consider each representation of the QME, and its stochastic Schr\"odinger equation~\cite{Belavkin1990,Carmichael1993,Wiseman2010}, which generates a piecewise-deterministic stochastic process (PDP)~\cite{Breuer_and_Petruccione}.  This encodes a probability distribution over quantum trajectories, which we call an \emph{ensemble} of quantum trajectories.  However, the relationship between representations and PDPs is not one-to-one: there are many representations that encode the same ensemble.
As a simple example, multiplying any jump operator by a phase factor does not change either the QME or the PDP.   Since the same process can be represented in many different ways, we refer to this as \emph{gauge invariance} of quantum trajectories and to the corresponding transformation as a \emph{gauge freedom}.  
In contrast to the well-established gauge freedoms of the QME~\cite{Avron2012}, gauge invariance of quantum trajectory dynamics has not been characterised before, to the best of our knowledge.

\noindent\emph{Contributions of this article} --We characterise these gauge freedoms by establishing necessary and sufficient conditions under which two representations of the quantum master operator lead to the same PDP. 
This is achieved by analysing the generators of the PDPs, which constitute \emph{unravelled QMEs}. In contrast to the QME, they govern the dynamics of probability distributions for pure density matrices, which represent conditional system states.  In fact, they fully determine the corresponding PDPs, and thus the ensembles of quantum trajectories.  Interestingly, the resulting gauge freedoms are much richer than simple multiplication of jump operators by phase factors or even their permutations; this is particularly apparent in systems with \emph{reset jumps}, which correspond to jump operators of rank 1, meaning that each jump operator resets the system's conditional state to a fixed destination.

As well as PDPs for the system state, we also consider \emph{labelled} quantum trajectories, which keep track of environmental measurement records.  We establish necessary and sufficient conditions for equivalence of these labelled quantum trajectory ensembles.  These gauge freedoms are weaker than those of the original PDP, reflecting that different representations may generate the same quantum trajectories, but distinct measurement records.
To further address this, we identify a coarse-graining procedure for measurement records, which yields \emph{partially-labelled} quantum trajectories with the same gauge freedoms as the PDP.

\noindent\emph{Implications} -- 
In the physical context, gauge invariance or equivalence of PDPs and quantum trajectory ensembles are relevant for weak symmetries of quantum master equations, and whether these are inherited by quantum trajectory ensembles. These issues are explored in  \cite{Usymm}, which relies extensively on the results presented here. When the goal is to generate quantum trajectory ensembles either in experiments or numerical simulations, our work clarifies the remaining gauge freedom that can be used to optimise these setups, e.g., by implementing a minimal equivalent representation.
Our results also reinforce the special status of reset jumps, which are common in many physical scenarios; for example energy level transitions in quantum optics experiments \cite{Sauter86,bergquist1986,nagourney1986}, superconducting circuits \cite{Weber_2016,minev19} and quantum dots \cite{jumps_dots,Tomm2024}. They also appear in the context of many-body quantum systems, e.g., when global projective measurements are performed at fixed rate \cite{Lami2024}. Finally, there has been much recent interest in a notion of `resetting' of quantum systems \cite{Mukherjee_2018,perfetto22,Kulkarni2023}, which when implemented at a constant rate can be encoded into the standard master equation formalism using necessarily global reset jumps.

The paper is structured as follows. In Sec. \ref{sec:background} we review the QME and  unravelled QMEs which prescribe stochastic quantum trajectories corresponding to different representations of the former. In Sec. \ref{sec:theorem_statement} we give the conditions for the unravelled QMEs for different representations to be the same and the resulting implications. Sec. \ref{sec:label_generators} discusses labelled and partially-labelled quantum trajectories, and the equivalence  conditions for associated stochastic processes. Sec.~\ref{sec:example} presents an illustrative example.  Sections \ref{sec:proof_1} and \ref{sec:labelled_gen_proof} detail the derivation of the results given in the sections \ref{sec:theorem_statement} and \ref{sec:label_generators} respectively. We conclude in Sec.~\ref{sec:conclusion}.

\section{System State Dynamics}\label{sec:background}

This section reviews the QME description of open quantum systems, and their unravelling as PDPs~\cite{Breuer_and_Petruccione}.

\subsection{Quantum master equation}
\label{sec:QME}

A Markovian open quantum system is governed by the quantum master equation (QME) \cite{Lindblad1976,Gorini1976,davies1974}
for the density matrix $\rho_t$:
\begin{equation}\label{eq:QME}
    \frac{d}{dt}\rho_t =
        \mathcal{L}(\rho_t)
\end{equation}
with
\beq
\label{eq:liouv}
{\cal L}(\rho_t) \equiv -i[H,\rho_t] + \sum_{k=1}^d\left(J_k\rho_tJ_k^\dag-\frac{1}{2}\{J_k^\dag J_k, \rho_t\}\right)
\eeq
where $H$ is the system \emph{Hamiltonian} and the $J_1,\dots,J_d$ are \emph{jump operators}, which describe the interaction of the system with the environment;
also $[A,B]=AB-BA$ denotes the commutator and $\{A,B\}=AB+BA$ the anti-commutator. 
The density matrix $\rho_t$ is the averaged system state and evolves deterministically as in Eq.~\eqref{eq:QME}, in contrast to the conditional state which evolves stochastically, see below.
We refer to the linear operator $\mathcal{L}$ as the \emph{quantum master operator}.
Note that $H$ and $J_k$ are operators acting on the system's Hilbert space while ${\cal L}$ is a super-operator, which acts on density matrices.

In general, there are many choices of Hamiltonian and jump operators that lead to the same quantum master operator $\mathcal{L}$.  Any specific choice for
\begin{equation}
    H, J_1, \dots, J_d
    \label{eq:rep-HJ}
\end{equation}
is called a \emph{representation} of the quantum master operator (which includes $d$ jump operators in this case). It is assumed that all $J_j\neq 0$.

 \emph{Minimal} representations of ${\cal L}$ have the smallest possible $d$, which we denote $d'$.   We distinguish the operators for such representations with primes.  Then, \emph{gauge invariance} of the QME can be summarised as follows~\cite{Breuer_and_Petruccione,Wolf2012}:
Given a minimal representation $H', J_1', \dots, J_{d'}'$, all other
representations 
$H, J_1, \dots, J_d$
of the same quantum master operator can be constructed as
\begin{subequations}\label{eq:QME_transformations}
\begin{align}
H & = H' + r\mathbb{1} - \frac{i}{2}\sum_{k=1}^{d'}\left(c_k^* J_k' - c_k J'^\dag_k\right),
\\
    J_j & = \sum_{k=1}^{d'} \mathbf{V}_{jk}\left(J_k'+c_k\mathbb{1}\right) \quad \text{for}\quad j\in\{1,2,\dots, d\} ,
\end{align}
\end{subequations}
where $d\geq {d'}$, $c_k \in \mathbb{C}$, $r\in \mathbb{R}$, and the matrix $\mathbf{V}\in \mathbb{C}^{ d\times d'}$ is an isometry, $\mathbf{V}^\dag \mathbf{V}=\mathbb{1}$.

\subsection{Quantum trajectories}\label{sec:traj}

We now turn to quantum trajectories \cite{Belavkin1990,Dalibard1992,Dum1992,Carmichael1993,Molmer93,Plenio1998,Wiseman2010,Daley2014,Landi2024}.
That is, we consider the stochastic evolution of a (pure) density matrix $\psi_t$ which represents the system state conditioned on a record of stochastic actions of jump operators on the system.  We refer to that construction as an \emph{unravelling} and the corresponding dynamics as the \emph{unravelled quantum dynamics}. The ensemble of quantum trajectories depends on the unravelling, via the jump operators and Hamiltonian.

 To motivate this stochastic construction in a physical setting, we further associate the action of each jump operator with the emission of an energy quantum from the system, which can be detected in the environment. Then, unravelling for a given representation encoded as in Eq.~\eqref{eq:rep-HJ} corresponds to a counting measurement scheme in which each action of a jump operator $J_k$ is associated with emission of a quantum of type $k$ that is detected in the environment.
The resulting $\psi_t$ follows a stochastic Schr\"odinger equation (SSE), which is the Belavkin equation~\cite{Belavkin1990}:
\begin{equation}\label{eq:Belavkin}
    d\psi_t = \mathcal{B}(\psi_t)dt + \sum_{k=1}^{d}\left\{\frac{\mathcal{J}_k(\psi_t)}{\Tr[\mathcal{J}_k(\psi_t)]} - \psi_t\right\}dq_{k, t},
\end{equation}
where $d\psi_t$ is the increment of $\psi_t$ in the interval $[t, t+dt]$ and
\begin{equation}\label{eq:B}
    \mathcal{B}(\psi) = -i\He\psi+i\psi\He^\dag - \psi\Tr(-i\He\psi+i\psi\He^\dag)
\end{equation}
with
\begin{equation}\label{eq:Heff}
    H_\text{eff} = H - \frac{i}{2}\sum_{k=1}^d J_k^\dag J_k,
\end{equation}
and
\begin{equation}\label{eq:superJ}
    \mathcal{J}_k(\psi)=J_k\psi J_k^\dag \,.
\end{equation}
In Eq.~\eqref{eq:Belavkin}, the conditional state changes either deterministically, with $dt$,  or due to random noise increments, $dq_{k, t}$, which take values $0$ or $1$ with the average $\mathbb{E}[dq_{k, t}] = \Tr[\mathcal{J}_k(\psi_t)]dt$. In the physical setting, $\mathcal{B}(\psi_t)dt$ is the change in the conditional state when no quanta are detected, while  $dq_{k, t}$ stands for the number of quanta of type $k$ detected  between times $t$ and $t+dt$.

The stochastic process for the state $\psi_t$ is a PDP~\cite{Breuer_and_Petruccione, Gardiner2004}: the state evolves by a continuous deterministic flow, punctuated by stochastic transitions. Quantum trajectories are sample paths  of this process, which we denote up to time $t$ as
\begin{equation}
\psi_{[0,t)}=	(\psi_{\tau})_{\tau \in [0,t)}. 
	\end{equation}
The deterministic flow is due to $\mathcal{B}(\psi_t)$, which we will refer to as the \emph{drift}.
Stochastic transitions, which we will call \emph{jumps}, are facilitated by the action of jump operators. If the state at a given time is $\psi$, then jumps facilitated by operator $J_k$ occur with the rate
\beq\label{eq:r_k}
r_k(\psi) = \Tr[\mathcal{J}_k(\psi)] \; ;
\eeq
 the associated conditional state changes from $\psi$ to
\beq\label{eq:D_k}
\mathcal{D}_k(\psi) =  \frac{\mathcal{J}_k (\psi)}{\Tr[\mathcal{J}_k (\psi)]}\,,
\eeq
which we refer to as the jump \emph{destination}. If ${\cal J}_k(\psi)=0$ then we set $\mathcal{D}_k(\psi)=0$.\footnote{If ${\cal J}_k(\psi)=0$ the jump rate is zero so such jumps never occur, and the value of $\mathcal{D}_k(\psi)$ in this case is purely conventional.} 

This PDP corresponds to the \emph{unravelled quantum master equation}~\cite{Breuer_and_Petruccione,Carollo2019,Carollo2021}
\begin{equation}\label{eq:UQME}
    \frac{\partial}{\partial t}P(\psi,t) = \mathcal{W}^\dag P(\psi,t),
\end{equation}
where $P(\psi,t)$ is the time-dependent probability distribution of the conditional state.
The generator $\mathcal{W}^\dag$ acts as 
\begin{multline}
\label{eq:Wdag}
 \mathcal{W}^\dag P(\psi,t) \equiv  -\nabla\cdot\left[\mathcal{B}(\psi)P(\psi,t)\right]
 \\ + \sum_{k=1}^{d} \int d\psi' \left[P(\psi',t)w_k(\psi',\psi)-P(\psi,t)w_k(\psi,\psi')\right] \,,
\end{multline}
where
\begin{align}
    w_k(\psi,\psi') &= \delta\!\left[ \psi' - \mathcal{D}_k (\psi)\right] r_k (\psi)
    \notag\\
    &= \delta\!\left\{ \psi' - \frac{\mathcal{J}_k (\psi)}{\Tr[\mathcal{J}_k (\psi)]}\right\}\Tr[\mathcal{J}_k (\psi)]
    \label{eq:w_k}
\end{align}
is the transition rate facilitated by jump operator of type $k$,  i.e., the  rate for jump of type $k$,  from current state $\psi$ into state $\psi'$.
In Eq.~\eqref{eq:Wdag}, we introduced a gradient, denoted $\nabla$.  The action of this gradient on scalar functions $f(\psi)$ gives the matrix with elements $(\nabla f)_{ab} = \partial f/\partial \psi_{ab}$. The divergence of a matrix $M(\psi)$ is therefore ${\nabla\cdot M = \sum_{ab}(\partial M_{ab}/\partial\psi_{ab})}$ \cite{Carollo2021}.
{The integration runs over all Hermitian matrices, see Appendix 1 of \cite{Carollo2019} for details.}

It will be convenient in the following to consider the adjoint operator $\mathcal{W}$ which is defined as $\langle f, \mathcal{W}^\dag g\rangle = \langle \mathcal{W} f,g\rangle$ where the inner product is $\langle f, g \rangle = \int f(\psi) g(\psi) d\psi$ for real functions $f(\psi)$ and $g(\psi)$.
This ${\cal W}$ is the (backwards) generator on the space of functions,
\begin{equation}\label{eq:W}
    \mathcal{W}f(\psi) = \mathcal{B}(\psi)\cdot\nabla f(\psi)+\sum_{k=1}^d\int d\psi' w_k(\psi,\psi')\left[f(\psi')-f(\psi)\right],
\end{equation}
{which in this work we refer to as the \emph{unravelled generator}.}
Expectation values evolve in time as $\frac{d}{dt} \mathbb{E}[ f(\psi_t) ] = \mathbb{E}\left[\mathcal{W}f(\psi_t)\right]$.
In particular, let initial conditions be such that $\mathbb{E}[\psi_0]=\rho_0$.  Then
\begin{equation}
\label{eq:Epsi-rho}
	\mathbb{E}[\psi_t]=\int d\psi P(\psi,t) \,\psi = \rho_t \, ,
\end{equation}
where $\rho_t$ is the average state of the system (in general a mixed state).
Taking the time derivative of this equation and using Eq.~\eqref{eq:W} with $f(\psi)=\psi$ 
shows that $\rho_t$ indeed obeys the QME in Eq.~\eqref{eq:QME}.\footnote{This $f(\psi)$ is a matrix valued function; in such cases $\mathcal{W}$ acts separately on each matrix element.}

We recall from Eq.~\eqref{eq:QME_transformations} that there are many representations (choices of $H, J_1, \dots, J_d$) that leave Eq.~\eqref{eq:QME} invariant: these are the gauge freedoms of the QME, which consequently preserve the evolution of the average state $\rho_t$.
 The PDP dynamics in Eq.~\eqref{eq:Belavkin} and the associated unravelled QME in Eq.~\eqref{eq:UQME} are constructed for a specific representation of the quantum master operator and thus can be expected to depend on that choice.  Indeed, in the physical setting where quantum jumps are associated with quanta detected in the environment, different representations of the quantum master operator correspond to different measurement bases, which could result in different back-action on the system states, leading to distinct quantum trajectory ensembles. Nevertheless, the next Section establishes what transformations between representations leave  the unravelled generator invariant. In turn, those determine the gauge freedoms in both the average and stochastic unravelled quantum dynamics.

\section{Gauge Invariance for Quantum Trajectories}\label{sec:theorem_statement}

This Section formulates the gauge freedoms of the unravelled quantum dynamics. These gauge freedoms are found by identifying which representations of the QME lead to the same unravelled generator.
The results are stated here, with proofs given Sec.~\ref{sec:proof_1}.

We find that while the unravelled generator fixes the Hamiltonian up to a constant, there are freedoms of the jump operators that depend on their partitioning into sets, according to their common destinations.  To explain this, we introduce the relevant \emph{sets of jumps with equal destinations} (SJEDs).
This is followed by the presentation of our first main result: the sufficient and necessary conditions for a pair of representations to have the same unravelled generators, which depend on the associated SJEDs. 
Finally, this result is used to fully describe the gauge freedoms of the unravelled QME.

\subsection{SJED}
\label{sec:SJED}

Consider a representation $H, J_1, \dots, J_d$
of a quantum master operator.  To define them, we say that
two operators $J_k,J_{k'}$ are jumps of equal destination (JEDs) if and only if for every $\psi$ one of the following holds: either $\mathcal{D}_k(\psi)=\mathcal{D}_{k'}(\psi)$ or $\mathcal{D}_k(\psi)=0$ or $\mathcal{D}_{k'}(\psi)=0$.  Physically: JEDs have the same destination whenever their rates are non-zero.  We define SJEDs $S_\alpha$ as follows: the jump labels $\{1,2,\dots,d\}$ are partitioned into sets $S_1,S_2,\dots,S_{d_C}$ such that if $J_k,J_{k'}$ are JEDs then they belong to the same set $S_\alpha$. (We use Greek indices for SJEDs to distinguish them from jump operators, which are indexed by Roman indices.)

The SJED definition is equivalent to
\beq
\label{eq:def-cgs}
k,k' \in S_\alpha \quad \Leftrightarrow \quad \forall {|\psi\rangle} \;\; \exists\, {c,c'}\; \; c\, J_k |\psi\rangle = c' J_{k'} |\psi\rangle , \;\; 
\eeq
where $c,c'\in\mathbb{C}$ depend in general on $k,k',|\psi\rangle$; in particular, either $c$ or $c'$ may be zero [if $\mathcal{D}_k(\psi)=0$ or $\mathcal{D}_{k'}(\psi)=0$].
With slight abuse of notation, the term SJED will be used interchangeably in the following for $S_\alpha$ (as above) and for the corresponding set of jump operators $\{ J_k \}_{k\in S_\alpha}$.   (The definition of JEDs is an equivalence relation between jump operators, and the SJEDs are the corresponding equivalence classes.)

To see why SJEDs are useful, we define a super-operator ${\cal A}_\alpha$ that describes
the composite action of SJED $\alpha$: 
\begin{equation}\label{eq:superA}
	\mathcal{A}_\alpha(\psi) = \sum_{k\in S_\alpha} \mathcal{J}_k(\psi) \; . 
\end{equation}
It follows from Eq.~\eqref{eq:def-cgs} that for $k\in S_\alpha$, either  $\mathcal{D}_k(\psi)= 0$ or 
\begin{equation}\label{eq:SJED_jump}
	\mathcal{D}_k(\psi) = \frac{\mathcal{A}_\alpha(\psi)}{\Tr[\mathcal{A}_\alpha(\psi)]}\,.
\end{equation}
In fact, the SJEDs are the maximal sets of jump operators with this property.\footnote{Indeed, ${\cal A}_\alpha(\psi)$ has rank $\leq1$ for all $\psi$  but when  $\alpha'\neq\alpha$ there always exists $\psi$  such that ${\cal A}_\alpha(\psi) + {\cal A}_{\alpha'}(\psi)$ has rank $\geq2$, i.e., the normalised state would be mixed due to differing destinations facilitated by those SJEDs for $\psi$.}
Hence,  [cf.~Eq.~\eqref{eq:w_k}]
\beq
\sum_{k\in S_\alpha} w_k(\psi,\psi')
= \delta\!\left\{ \psi' - \frac{\mathcal{A}_\alpha (\psi)}{\Tr[\mathcal{A}_\alpha (\psi)]}\right\} \Tr[\mathcal{A}_\alpha (\psi)] .
\label{equ:wA-lemma}
\eeq
That is, the rates of stochastic jumps facilitated by operators from the same SJED can be naturally grouped together.  Moreover, the generator of the unravelled quantum dynamics in Eq.~\eqref{eq:Wdag} only depends on the summed rates in Eq.~\eqref{equ:wA-lemma}, which are naturally expressed in terms of the composite jump action operators defined in Eq.~\eqref{eq:superA}.  Since jump operators in different representations may still lead to the same composite actions for their SJED, they allow for gauge freedom in the unravelled quantum dynamics to remain.

We now show that SJEDs can be separated into two distinct types: 
\begin{itemize}
\item \textbf{Reset SJED}: all jump operators in the SJED  are of the form
\begin{equation}\label{eq:reset}
	J_k = \sqrt{\gamma_k}|\rs_\alpha\rangle\!\langle\xi_k| \, \text{ for } k\in S_\alpha,
\end{equation}
    where $|\chi_\alpha\rangle$ is the same for all jump operators in the SJED, but $\gamma_k\in\mathbb{R}$ and $|\xi_k\rangle$ in general depend on $k$; also $\langle\chi_\alpha|\chi_\alpha\rangle=1=\langle \xi_k|\xi_k\rangle$. 
    Hence, for $k\in S_\alpha$ and taking $\psi$ with non-zero jump rate [${\cal D}_k(\psi)\neq0$], the operator ${\cal J}_k$ always
        \emph{resets} the conditional state to the same fixed destination  $|\chi_\alpha\rangle\langle \chi_\alpha|$.  Moreover,
    \begin{equation}
        \mathcal{A}_\alpha(\psi) = |\chi_\alpha\rangle\!\langle\chi_\alpha| \,
        \text{Tr}(\Gamma_\alpha\psi)
                \label{eq:A-gamma}
    \end{equation}
    with
        $\Gamma_\alpha = \sum_{k\in S_\alpha} \gamma_k |\xi_k\rangle\!\langle\xi_k|$. 
        \item \textbf{Non-reset SJED}: all jump operators in the SJED are proportional to a single operator $J^{(\alpha)}$:
    \begin{equation}\label{eq:non_reset}
          J_k = \lambda_k J^{(\alpha)} \text{ for } k\in S_\alpha
    \end{equation}    where $\lambda_k\in\mathbb{C}$ and $J^{(\alpha)}$ has rank $>1$ (otherwise this is a reset SJED); also $\mathrm{Tr}[J^{(\alpha)\dagger} J^{(\alpha)}]=1$. The resulting composite action is proportional to the action of the single jump operator,
        \begin{equation}
    	\mathcal{A}_\alpha(\psi) = |\lambda^{(\alpha)}|^2\,	\mathcal{J}^{(\alpha)}(\psi)
    	\label{eq:A-lambda}
    \end{equation}
    with
    $\lambda^{(\alpha)} = \sqrt{\sum_{k\in S_\alpha} |\lambda_k|^2} .$
\end{itemize}

The fact that both reset and non-reset SJEDs obey~\eqref{eq:def-cgs} can be verified directly from their definitions. The proof that these are the only possibilities is given in Appendix~\ref{app:sjed}.

From a physical perspective, reset SJEDs are interesting because rank-1 jump operators appear in many physical settings, as discussed in Sec.~\ref{sec:intro}.  Furthermore,  for dynamics with jump operators of this type only, a semi-Markov mapping of the unravelled dynamics exists~\cite{Carollo2019,Carollo2021,Brown24}, which simplifies the sampled space in stochastic simulations.  Non-reset SJEDs can be relevant in experimental settings if the environment consists of multiple identical reservoirs or classical noise is present in the measurement process. In simulations it would be natural to exploit the associated gauge freedom and combine them into a single jump operator from the start.

\subsection{Equality of unravelled generators}

We are now ready state our first main theorem.
Given two representations of the same quantum master operator, 
\begin{equation}
	H, J_1, \dots, J_d\quad\text{and}\quad \tilde H, \tilde J_1, \dots, \tilde J_{\tilde d}\,,
\end{equation}
the corresponding SJEDs are denoted as $S_1, \dots, S_{\dc}$
and $\tilde S_1, \dots, \tilde S_{\dct}$,
and give rise to the associated super-operators for the composite action of their jump operators [cf.~Eq.~\eqref{eq:superA}]
\begin{equation}
\mathcal{A}_1, \dots, \mathcal{A}_{\dc}\quad\text{and}\quad \tilde{\mathcal{A}}_1, \dots, \tilde{\mathcal{A}}_{\tilde{d}_c}.
\end{equation}
The numbers of SJEDs are $\dc\leq d$ and $\dct\leq \tilde{d}$. 
Then the conditions for these two representations to describe the same PDP are given by the following Theorem.

\vspace{4mm}
\noindent
$\bullet \;$ \textbf{Theorem~1:}

\noindent
\emph{For two representations of a given quantum master operator, $H, J_1, \dots, J_d$
and
$\tilde H, \tilde J_1, \dots, \tilde J_{\tilde d}$,
the corresponding unravelled generators obey
\begin{equation}\label{eq:W=Wtilde}
    \tilde{\mathcal{W}}=\mathcal{W},
\end{equation}
if and only if
}
\begin{subequations}\label{eq:theorem_rep}
\begin{align}
&        \Tilde{H}  = H + r\mathbb{1}, \quad r\in\mathbb{R},
 \label{eq:theorem_cond_H}
\\
 &      \dct  = \dc \quad\text{and}\quad \Tilde{\cgo}_{\alpha} = \cgo_{\pic(\alpha)} \;\; \forall \alpha, \; 
\label{eq:theorem_cond_J}
\end{align}
\end{subequations}
\emph{for some permutation $\pic$ of $\{1,2,\dots,\dc\}$}.
\smallskip

Note that the definition of SJED ensures that the permutation $\pi$ appearing in Theorem $1$ is uniquely defined (it is not possible that ${\cal A}_\alpha={\cal A}_\beta$ for $\alpha\neq\beta$). The fact that Eq.~\eqref{eq:theorem_rep} is sufficient for Eq.~\eqref{eq:W=Wtilde} can be verified directly from the definition of ${\cal W}$, with the aid of Eq.~\eqref{equ:wA-lemma}, as we now explain.  Showing the converse requires more work, this proof is given in Sec.~\ref{sec:proof_1}. {The proof shows that the algebraic condition on jump operators Eq.~\eqref{eq:theorem_cond_J} is sufficient to ensure that the stochastic jumps of the two representations occur between the same quantum states, with the same rates.  This result has potential relevance more generally in stochastic processes, beyond the specific case of quantum trajectories.}

To see that Eq.~\eqref{eq:theorem_rep}  implies  Eq.~\eqref{eq:W=Wtilde}, note first that Eq.~\eqref{eq:theorem_cond_J} requires $\{{\cal A}_\alpha\}_{\alpha=1}^{\dc}$ and  $\{\tilde{\cal A}_{\alpha}\}_{\alpha=1}^{\tilde{d}_C}$ to be equal as sets (which equality allows for any permutation of their elements). This  ensures equal rates for stochastic transitions in both representations [cf.~Eqs.~\eqref{eq:w_k} and~\eqref{equ:wA-lemma}].  It also ensures that the anti-Hermitian parts of the effective Hamiltonian are equal between the two representations [Eq.~\eqref{eq:Heff}].  Then, Eq.~\eqref{eq:theorem_cond_H} implies that the Hermitian parts of the effective Hamiltonians are equal up to a constant.  Together, these facts ensure that the drift operators [Eq.~\eqref{eq:B}] are  equal for the two representations.

In other words, two different representations have the same unravelled generator if their Hamiltonians are the same up to a constant (which gives rise to a global phase for a state vector, but does not change the corresponding density matrix) and their SJEDs act in the same way (so that the conditional state is changed identically and with the same overall rates).

An interesting special case of Theorem~1 is  $\dc= d$ and $\dct=\tilde{d}$ so that
each SJED contains a single jump operator.  Then the condition in Eq.~\eqref{eq:theorem_cond_J} implies that jump operators in the two representations are related simply by a permutation ($\pi$) up to a relative phase ($\phi_k\in \mathbb{R}$),
\begin{equation}
    \tilde{J}_k = {\rm e}^{i\phi_k} J_{\pi(k)}  \quad \forall\, k.
    \label{eq:single-J-phi}
\end{equation}
This case is considered explicitly in the proof in Sec.~\ref{sec:proof_1}. In general, while Eq.~\eqref{eq:theorem_cond_J}  constrains the composite action of SJEDs, it leaves more freedom in the choice of jump operators, compared with Eq.~\eqref{eq:single-J-phi} (see Sec.~\ref{sec:gauge-cgs}, below).

An important corollary of Theorem~1 is that $\mathcal{W}^\dag = \tilde{\mathcal{W}}^\dag$ under the conditions in Eq.~\eqref{eq:theorem_rep}, which means that their stationary distributions over conditional states are identical, and also $P(\psi,t)$ at any time $t$ is the same for both representations provided that one considers the same initial distribution.
The two different representations actually produce the same ensemble of quantum trajectories in that case [but Eq.~\eqref{eq:theorem_rep} does not guarantee equivalence of their measurement records, see Sec.~\ref{sec:label_generators}]. We clarify what transformations between representations allow for that invariance next.

Finally, observe that if two representations give rise  
to different QMEs, then the relationship in Eq.~\eqref{eq:Epsi-rho} between the QME and unravelled dynamics means that the unravelled generators do not coincide, ${\cal W}\neq\tilde{\cal W}$, so at least one of the conditions in Theorem~1 must be violated.

\subsection{Gauge freedoms of unravelled quantum dynamics}
\label{sec:gauge-cgs}

Given Theorem~1, a natural question follows: which representations give rise to a given a set of super-operators $\{{\cal A_\alpha}\}_{\alpha=1}^{d_C}$, and hence to the same unravelled dynamics? 
To answer this, we note that equality of $ \Tilde{\cgo}_{\alpha}$ and $\cgo_{\pic(\alpha)} $ in Eq.~\eqref{eq:theorem_cond_J} is the  condition [cf.~Eq.~\eqref{eq:superA}]
\beq\label{eq:superAJ}
\sum_{j\in \tilde S_\alpha} \tilde{\mathcal{J}}_{j}  = \sum_{k\in S_{\pic(\alpha)}} {\mathcal {J}}_{k} \,.
\eeq
From Eq.~\eqref{eq:superJ}, it then follows that the sets $\{\tilde J_{j}\}_{j\in \tilde S_{\alpha}}$  
and $\{J_k\}_{k\in S_{\pic(\alpha)}}$  are different representations of the same completely positive super-operator, $ \Tilde{\cgo}_{\alpha}=\cgo_{\pic(\alpha)} $, and thus are related by an isometry~\cite{Wolf2012}.  Appendix~\ref{app:gauge-indiv} describes the 
gauge freedoms associated with the composite action operator for each SJED.

Using these results, we now describe the full gauge freedom of the unravelled quantum dynamics,
see Appendix~\ref{app:gauge-comb} for details.
It is convenient to consider a minimal representation (in the sense that each ${\cal A}_\alpha$ is represented by a minimal number of jump operators, see Appendix~\ref{app:gauge-sjed}). 
Then the gauge freedoms of the unravelled dynamics are given as follows:

\bigskip
\noindent\emph{Suppose that $H', J_1',\dots, J_{d'}'$ is a representation of a given quantum master operator in which all SJED actions ${\cal A}'_1,\dots,{\cal A}'_{\dc'}$ have minimal representations.  
Then
$H, J_1, \dots, J_d$
has the same unravelled generator if and only if $\dc'=\dc$ and there exists a permutation $\pic$ of $\{1,\dots,\dc\}$ and a matrix $\mathbf{V}\in \mathbb{C}^{d \times d'}$ such that
\begin{subequations}\label{eq:V_rep_connection}
\begin{align}
&        H  = H' + r\mathbb{1}, \quad r\in\mathbb{R},
\label{equ:H-V-rep}
\\
&	J_j = \sum_{k=1}^{d'} \mathbf{V}_{jk} J_k'\,,
\label{equ:V-V-rep}
\end{align}
where $\mathbf{V} = \sum_{\alpha} \mathbf{V}^{(\alpha)}$ with
\begin{equation}
	\mathbf{V}_{jk}^{(\alpha)} = 0 \quad \text{unless} \quad  j\in  S_\alpha, \, k\in S'_{\pic(\alpha)}
	\label{equ:pad-V-rep}
\end{equation}
and 
 \begin{equation}
 \sum_{j\in  S_\alpha} [\mathbf{V}^{(\alpha)}_{jk}]^*\mathbf{V}^{(\alpha)}_{jk'}=\delta_{kk'}  \quad\text{for}\quad k,k'\in S'_{\pic(\alpha)} \; .
 \label{eq:V-big-isom}
 \end{equation}
\end{subequations}
}
\smallskip

\noindent This result is analogous to the characterisation of the gauge freedoms in the QME given in Eq.~\eqref{eq:QME_transformations}.
It can be verified from Eq.~\eqref{eq:V-big-isom} that $\mathbf{V}$ is indeed an isometry, that is, $\mathbf{V}^\dag \mathbf{V}=\mathbb{1}$ (see Appendix~\ref{app:gauge-sjed}).  Hence Eq.~\eqref{eq:V_rep_connection} implies Eq.~\eqref{eq:QME_transformations}. Indeed this must be the case because ${\cal W}=\tilde{\cal W}$ implies that the two representations have the same QME.  However, the transformations in Eq.~\eqref{eq:V_rep_connection} are more constrained than those in Eq.~\eqref{eq:QME_transformations}, due to Eq.~\eqref{equ:pad-V-rep}.  That is, the unravelled QME has less gauge freedoms than the QME.

To end this Section, note that given any representation, one can always construct a representation $H', J_1',\dots, J_{d'}'$ which all the SJED actions ${\cal A}_1,\dots,{\cal A}_{\dc}$ have minimal representations and Eq.~\eqref{eq:W=Wtilde} holds.  From this latter representation, one can apply the gauge freedoms of Eq.~\eqref{eq:V_rep_connection} to construct all possible representations of the resulting unravelled dynamics (including the representation already given).

\section{Gauge equivalence\\for labelled quantum trajectories}\label{sec:label_generators}

As explained in Sec.~\ref{sec:background}, the conditional state of the system evolves by a PDP which consists of deterministic segments, 
punctuated by jumps. When those jumps are associated with emissions of quanta (for example, photons) that can be detected in the environment, such detection events can be collected in a measurement record. 
This section derives conditions under which two representations of a QME share the same ensembles of quantum trajectories and measurement records (up to permutations of the latter).

\subsection{Labelled quantum trajectories}\label{sec:WF}

Recall that each action of a jump operator $J_k$ is associated with emission of a quantum of type $k$.
Writing $q_{k,t}$ for the number of quanta of type $k$ emitted between times 0 and $t$, the random noise $dq_{k,t}$ in Eq.~\eqref{eq:Belavkin} is simply the increment of $q_{k,t}$ at time $t$. Therefore, the evolution of $\bm{q}_{t}=(q_{1,t},\dots,q_{d,t})$  encodes a measurement record that includes the types of all emitted quanta, and the times at which they were emitted.  
%
A sample path of the dynamics for the conditional state $\psi_t$ and the measurement counts $\bm{q}_{t}$ is
\begin{equation}\label{eq:path_psi_q}
    (\psi_{[0,t)}, \bm q_{[0,t)}) = (\psi_\tau, \bm q_\tau)_{\tau\in [0,t)}\;,
\end{equation}
where  $\psi_t$ is the conditional system state at time $t$ as before. We  refer to Eq.~\eqref{eq:path_psi_q} as a \emph{labelled quantum trajectory} and the corresponding dynamics as \emph{labelled quantum dynamics}.
Indeed, jumps of $\psi_t$ are accompanied by transitions in $\bm{q}_t$, from which one may infer which jump operator facilitated the jump. 

The generator of the labelled quantum dynamics is denoted as $\WF$, it
 acts on functions
$f(\psi,\bm q)$, where $\bm q$ is a $d$-vector with integer entries, as [cf.~Eq.~\eqref{eq:W}]
\begin{equation}\label{eq:Wf}
    \WF f(\psi,\bm q) = {\cal B}(\psi)\cdot\nabla f(\psi,\bm q) 
    + \sum_{k=1}^{d} \int d\psi' \, w_k(\psi,\psi') [ f(\psi',\bm{q}+\bm{e}_k) - f(\psi,\bm{q}) ],
\end{equation}
where $\bm{e}_k$ is a $d$-vector with a single non-zero entry of 1 in the $k$th position [that is, $(\bm{e}_k)_j=\delta_{jk}$]. The subscript $F$ indicates that we consider full measurement records and in particular, the types of all quanta are recorded.

To formulate gauge freedoms of the labelled quantum trajectories, it is useful to consider measurement records with jump types related by permutations. (Such transformations do not affect the conditional state dynamics and are invertible in the classical sense of post-processing the records, so we regard trajectory ensembles with permuted jump types as being equivalent.) 
To this end, we define a permutation operation that acts on functions as
\begin{equation}\label{eq:Pi}
    \Pi^\dag f(\psi,\bm q) = f[\psi, \bm\pi(\bm q)]\,,
\end{equation}
where $\pi$ is a permutation of $\{1,2,\dots,d\}$ and the action of this permutation on a $d$-vector is defined as the corresponding permutation of its entries, i.e.,  $\bm\pi(\bm{q})=(q_{\pi(1)},q_{\pi(2)},\dots,q_{\pi(d)})$.
Then \emph{gauge equivalence} of the labelled quantum dynamics means that for two representations the corresponding generators are \emph{equivalent}: there exists a permutation $\pi$ such that
\beq\label{eq:PiWf0}
 \Pi \tilde{\mathcal{W}}_F \Pi^\dag = \mathcal{W}_F
\eeq
with $\Pi$ defined as in Eq.~\eqref{eq:Pi}. This equivalence  means that 
the labelled quantum trajectory ensembles for the two representations are identical up to permutation of $\bm{q}_t$ by $\pi$. [For comparison, we recall that the gauge invariance in the sense of Eq.~\eqref{eq:W=Wtilde}, implies that the (unlabelled) quantum trajectory ensembles are identical.]
Note that the existence of $\pi$ restricts both representations to have the same number of jump operators $d$.
The following theorem gives necessary and sufficient conditions for the gauge equivalence.
  
\vspace{4mm}
\noindent
$\bullet \;$ \textbf{Theorem~2:}

\noindent
\emph{Consider two representations of a given quantum master operator, $H, J_1, \dots, J_d$ and
\break
$\tilde H, \tilde J_1, \dots, \tilde J_{{d}}$, both of which have $d$ jump operators.  Given a permutation $\pi$ of $\{1,2,\dots,d\}$ and defining $\Pi$ as in Eq.~\eqref{eq:Pi}, the corresponding generators for the labelled quantum dynamics obey
\begin{equation}\label{eq:PiWf}
    \Pi \tilde{\mathcal{W}}_F \Pi^\dag = \mathcal{W}_F
\end{equation}
if and only if
\begin{equation}\label{eq:WF_cond}
    \Tilde{H}=H+r\mathbb{1}, \qquad \Tilde{J}_k = {\rm e}^{i\phi_k} J_{\pi(k)}  \; \forall k 
\end{equation}
for some $r\in\mathbb{R}$ and $\phi_k\in\mathbb{R}$.
}
\smallskip

This theorem is proved in Sec.~\ref{sec:WF_proof}.   Here, we note that the condition in Eq.~\eqref{eq:WF_cond} implies Eq.~\eqref{eq:theorem_rep},  so using Theorem~1 we obtain that the ensembles of quantum trajectories are equal for any two representations for which Theorem~2 holds.  However, the condition in Eq.~\eqref{eq:WF_cond} is more restrictive, because Eq.~\eqref{eq:PiWf} implies that the two representations lead to the joint dynamics of quantum trajectories and measurement records being \emph{equivalent}, i.e., identical up to the given permutation $\pi$ of the records.

Indeed, the gauge freedoms in Eq.~\eqref{eq:V_rep_connection} -- which follow from Theorem~1 -- allow for all  isometric transformations of jump operators within SJEDs.  On the other hand, the gauge freedoms following from Theorem 2 are as follows:
 [cf.~Eq.~\eqref{eq:QME_transformations}]:

\bigskip
\noindent\emph{For two representations  $H, J_1,\dots, J_d$ and 	$\tilde{H}, \tilde J_1, \dots, \tilde J_{\tilde d}$ of a given quantum master operator, the corresponding generators of the labelled quantum dynamics are equivalent if and only if $d=\tilde d$ and there exists a permutation $\pi$ of $\{1,\dots,d\}$  and a matrix $\mathbf{V}\in \mathbb{C}^{d \times d}$ such that
{\begin{subequations}\label{eq:V_rep_connection2}
	\begin{align}
		&        \Tilde{H}  = H + r\mathbb{1}, \quad r\in\mathbb{R},
		\label{equ:H-V-rep2}
		\\
		&	\tilde J_j = \sum_{k=1}^d \mathbf{V}_{jk} J_k\,,
		\label{equ:V-V-rep2}
	\end{align}
	where
	\begin{equation}
		\mathbf{V}_{jk} = {\rm e}^{i\phi_j} \delta_{\pi(j) k }, \quad \phi_j\in\mathbb{R}.  
		\label{eq:V-big-isom2}
	\end{equation}
\end{subequations}
}}
\smallskip

Recalling Eq.~\eqref{eq:single-J-phi}, one sees that the gauge freedoms of unravelled quantum dynamics and the gauge equivalence of labelled quantum dynamics coincide when every SJED contains exactly one jump.  In general, Eq.~\eqref{eq:V_rep_connection2} requires the related jump operators from two representations to have the same rate 
[cf.~Eq.~\eqref{eq:WF_cond}].  
Note however that the permutation $\pi$ in Eq.~\eqref{eq:V_rep_connection2} does not need to be unique: If (and only if)  a given representation has multiple jump operators that are equal up to a phase, then their permutation can be composed with $\pi$, to obtain an alternative representation.
In that case Theorem~2 holds simultaneously for different permutations; an example is given in Sec.~\ref{sec:example}.

Finally, we note that that if we require equality of labelled generators ($\tilde{\mathcal{W}}_F= \mathcal{W}_F$) instead of equivalence as in Eq.~\eqref{eq:PiWf0}, the ensembles of labelled quantum trajectories are identical, but the only remaining gauge freedom is that of shifting the Hamiltonian by a real constant and multiplying the jump operators by phases. This follows directly by considering the trivial permutation [$\pi(k)=k$] in Theorem~2. Such equality can be obtained for any two equivalent generators by relabelling jump operators accordingly to the permutation in Eq.~\eqref{eq:WF_cond}.

\subsection{Partially-labelled quantum trajectories}\label{sec:WC}

We have seen that the equivalence of labelled quantum trajectory ensembles permits less gauge freedoms than the equality of (unlabelled) quantum trajectory ensembles.  To address this, we now construct partially-labelled quantum trajectories for which the equivalence allows \emph{the same} gauge freedoms as those of the (unlabelled) quantum trajectories. This construction clarifies what information about measurement records is already present in the unravelled quantum dynamics and further elucidates properties of pairs of representations for which Theorem~1 holds but Theorem~2 fails.

Our construction is based on \emph{coarse-graining} of  measurement records:
instead of recording the type of each emitted quantum, we only record the SJED to which the relevant jump operator belongs.
The corresponding SSE is then given by 
\begin{equation}\label{eq:Belavkin_CG}
    d\psi_t = \mathcal{B}(\psi_t)dt + \sum_\alpha\left\{\frac{\mathcal{A}_\alpha(\psi_t)}{\Tr[\mathcal{A}_\alpha(\psi_t)]} - \psi_t\right\}dQ_{\alpha, t},
\end{equation}
where $dQ_{\alpha,t}$ is an increment at time $t$ of $Q_{\alpha,t} = \sum_{k\in S_\alpha} q_{k,t}$, which is the number of jumps between times $0$ and $t$ that were facilitated by jump operators $J_k$ with $k\in S_\alpha$  [cf.~Eq.~\eqref{eq:Belavkin}].
A sample path for the dynamics of the conditional state $\psi_t$ and the coarse-grained measurement counts $\bm{Q}_t=(Q_{1,t},...,Q_{\dc,t} )$ is
\begin{equation}\label{eq:path_psi_Q}
    (\psi_{[0,t)}, \bm Q_{[0,t)}) = (\psi_\tau, \bm Q_{\tau})_{\tau\in[0,t)},
\end{equation}
which we call a \emph{partially-labelled quantum trajectory}. The corresponding dynamics is called \emph{partially-labelled quantum dynamics}.

The generator of partially-labelled quantum dynamics is denoted by $\WC$; it acts on functions $f(\psi,\bm{Q}) $ where $\bm{Q}$ is a $\dc$-vector with integer entries, as
\begin{multline}\label{eq:WC}
    \WC f(\psi,\bm{Q}) = {\cal B}(\psi)\cdot\nabla f(\psi,\bm{Q})  
    +  \sum_{\alpha=1}^{\dc} \int w^{(\alpha)}(\psi,\psi') [ f(\psi',\bm{Q}+\bm{E}_\alpha) - f(\psi,\bm{Q}) ] d\psi'\,,
\end{multline}
where
$w^{(\alpha)}(\psi,\psi') = \sum_{k\in \SJED}w_k(\psi,\psi')$,
cf.~Eq.~\eqref{equ:wA-lemma},
and $(\bm{E}_\alpha)_\beta = \delta_{\alpha\beta}$.  The subscript $C$ refers to the coarse-grained measurement records.

Similarly to the case of labelled quantum dynamics, we consider \emph{gauge equivalence} of partially-labelled quantum dynamics for two representations to hold when the corresponding generators are \emph{equivalent} with respect to some permutation operation [cf.~Eq.~\eqref{eq:Pi}]
\begin{equation}\label{eq:Pi_Q}
	\Pi_C^\dag f(\psi,\bm Q) = f[\psi, \bm\pic(\bm Q)]\,,
\end{equation}
where $\pic$ is a permutation  of $\{1,2,\dots,\dc\}$ that acts on $\bm Q$ by permuting its entries. That is, for the pair of equivalent generators, there exists $\pic$ such that [cf.~Eq.~\eqref{eq:PiWf0}]
\begin{equation}\label{eq:pi_tilde_Wc}
   \Pi_C \tilde{\mathcal{W}}_C \Pi_C^\dag=  \mathcal{W}_C,
\end{equation} 
which implies that $\dct=\dc$. 
We prove the following theorem in Sec.~\ref{sec:WC_proof}.

\vspace{4mm}
\noindent
$\bullet \;$ \textbf{Theorem~3:}

\noindent
\emph{
Consider two representations of a given quantum master operator, $H, J_1, \dots, J_d$ 
and
$\tilde H, \tilde J_1, \dots, \tilde J_{\tilde d}$ 
both of which have $\dc$ SJEDs.  Given a permutation $\pic$ of $\{1,2,\dots,\dc\}$ and taking $\Pi$ as in Eq.~\eqref{eq:Pi_Q},
the generators for the partially-labelled quantum trajectories obey
\begin{equation}
    \Pi_C \tilde{\mathcal{W}}_C\Pi^\dag_C = \mathcal{W}_C
    \label{eq:PiWc}
\end{equation}
if and only if
\begin{equation}\label{eq:Wc_cond}
    \Tilde{H} = H + r\mathbb{1}, 
    \qquad \Tilde{\cgo}_{\alpha} = \cgo_{\pic(\alpha)} \; \forall \alpha
\end{equation}
for some $r\in\mathbb{R}$.
}
\smallskip

Similar to Theorem~2, the condition in Eq.~\eqref{eq:PiWc} means that the ensembles of partially-labelled quantum trajectories for the two representations are identical up to the given permutation $\pic$ of coarse-grained measurement records. By the definition of the SJED, there can only be one permutation $\pi$ for which Theorem~3 holds. Nevertheless, the conditions in Eq.~\eqref{eq:Wc_cond} allow more freedom than those of Theorem~2 in Eq.~\eqref{eq:WF_cond}, as the requirement of coarse-grained measurement records being the same up to a given permutation is less stringent than the analogous requirement for full measurement records. In fact, the conditions of Theorem~3 are almost identical to conditions in Eq.~\eqref{eq:theorem_rep} of Theorem~1: the only difference is that Theorem~3 applies for a given permutation $\pi$ while Theorem~1 allows for any permutation.  That is, Theorem~1 holds if and only if there exists $\pi$ such that Theorem~3 holds.
This can be conveniently formulated in terms of gauge equivalence: the generators of the partially-labelled dynamics for two representations are equivalent if and only if Theorem~1 holds. Therefore, the gauge freedoms corresponding to the gauge equivalence for the partially-labelled quantum dynamics are described by Eq.~\eqref{eq:V_rep_connection}  [and include the gauge freedoms  described by Eq.~\eqref{eq:V_rep_connection2}]. 

It also follows that if two representations have the same ensembles of (unlabelled) quantum trajectories, their ensembles of partially-labelled quantum trajectories are \emph{equivalent}, i.e.,  the same up to some permutation of coarse-grained measurement records. In general, however, their ensembles of labelled quantum trajectories do not have to be equivalent, see the example in Sec.~\ref{sec:example}.

\newcommand{\0}{\downarrow\downarrow}
\newcommand{\1}{\uparrow\uparrow}
\newcommand{\2}{\uparrow\downarrow}
\newcommand{\3}{\downarrow\uparrow}

\section{Example}\label{sec:example}

Here, we present an example of different representations for the same QME that illustrate the conditions in Theorems~1,~2, and~3. The example features both reset and non-reset SJEDs, for which minimal representations are also discussed. We also illustrate the uniqueness or lack thereof for permutations appearing in these theorems. A second example is given in Appendix~\ref{app:example}.

\begin{figure}
    \centering
	\includegraphics[width=0.75\linewidth]{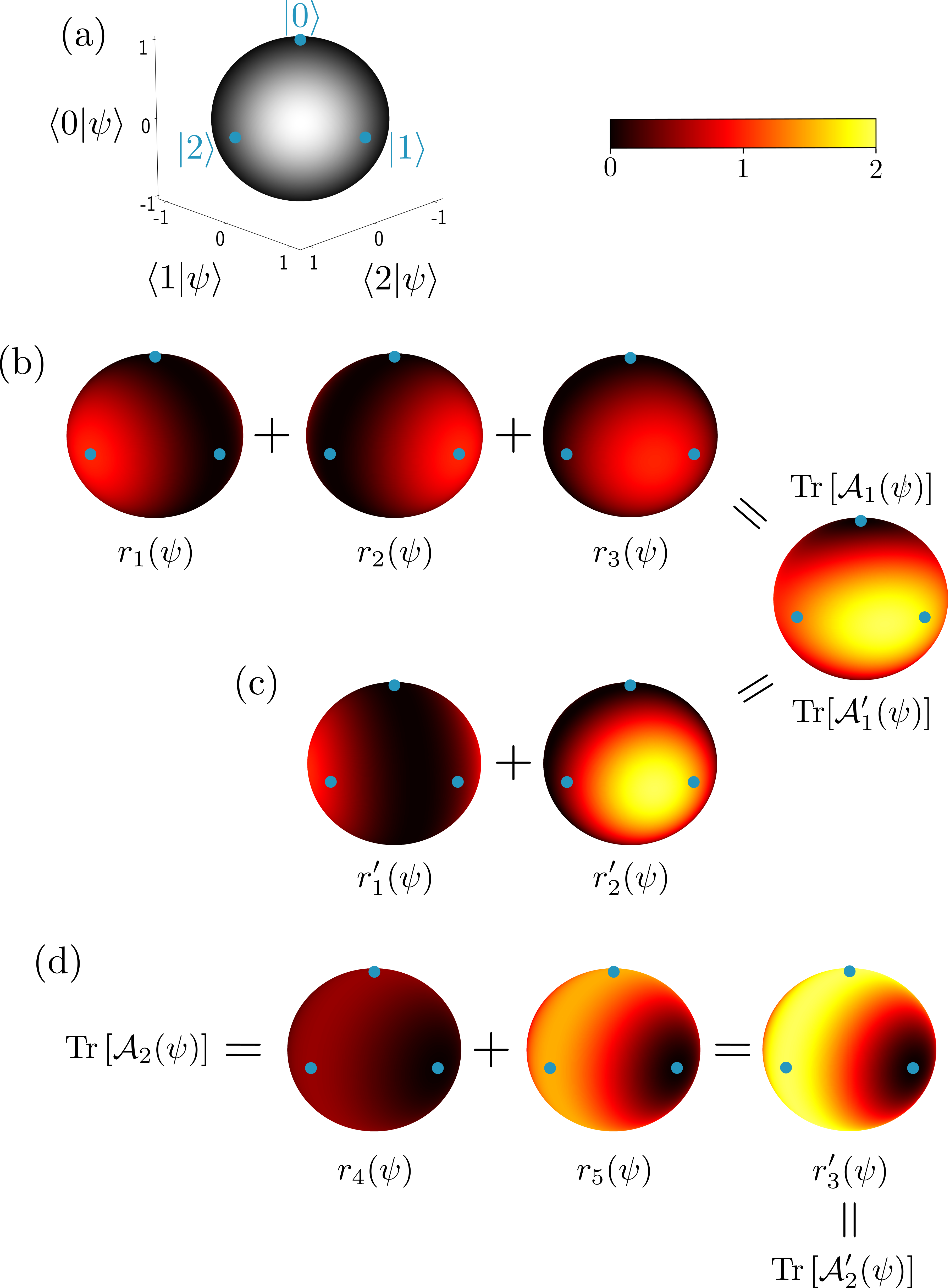}
	\caption{Jump rates in the example of Sec.~\ref{sec:example} with parameters $\theta=\pi/6, \, \gamma=1, \, \vartheta=\pi/3, \, \lambda=2$.
	\textbf{(a)} Pure states with real coefficients in the basis $|0\rangle$, $|1\rangle$, $|2\rangle$ can be represented as points on the surface of a sphere. The lighter grey shading indicates a shorter distance from the observer. \textbf{(b)} Rates for jump operators $J_{1}$, $J_{2}$, and $J_{3}$ in Eq.~\eqref{eq:ex1:rep1} together with their sum, which corresponds to the rate of their composite action $\Tr[{\cal A}_1(\psi)]=\Tr(\Gamma\psi)$. 
	\textbf{(c)} Rates for jump operators $J'_{1}$ and $J'_{2}$ in Eq.~\eqref{eq:ex1:rep2}, which yield the same composite action rate as in (a). 
	 \textbf{(d)} Rates for jump operators $J_4$ and $J_5$ in Eq.~\eqref{eq:ex1:rep1} and their sum $\Tr[\mathcal{A}_2(\psi)] = |\lambda|^2\mathcal{J}(\psi)$, which coincides with the rate for operator $J_3'$. Note that these jump rates are all proportional, $r_4(\psi)=r_5(\psi)/3=r_3'(\psi)/4$.}
\label{fig:jump_rate_rep}
\end{figure}

\subsection{Representations}

We consider a 3-level system (a qutrit) with the basis $|0\rangle$, $|1\rangle$, $|2\rangle$. We take an arbitrary Hamiltonian $H$, its form plays no role in the following, and we define jump operators
\begin{align}\label{eq:ex1:rep1}
    &J_1 = \sqrt{\gamma}|0\rangle\!\langle 1|, \quad
    J_2 =\sqrt{\gamma} |0\rangle\!\langle 2|, \quad
    J_3  =\sqrt{\gamma} |0\rangle\!\left(\cos\theta\langle 1| + \sin\theta\langle 2|\right), 
    \nonumber\\
    &J_c (\vartheta ) = \lambda \deph \cos\vartheta , \quad 
    J_s (\vartheta,\phi ) = {\rm e}^{i\phi} \lambda\deph \sin\vartheta ,
\end{align}
where $\deph = (|2\rangle\!\langle2|-|0\rangle\!\langle 0|)/\sqrt{2}$,
the parameter
$\theta\in \mathbb{R}$ is given, and $\vartheta,\phi\in \mathbb{R}$ can be varied, giving rise to different representations of the same QME (see below).  We assume $\vartheta\neq n\pi/2$ so that neither $J_c=0$ nor $J_s=0$.

The jump operators $J_1$, $J_2$,  $J_3$, are all of rank $1$, so their destinations are the same $\mathcal D_1(\psi)=\mathcal D_2(\psi)=\mathcal D_3(\psi)=|0\rangle\!\langle 0|$, except for the special cases where ${\cal D}_k(\psi)=0=r_k(\psi)$.   However, the rates $r_1(\psi)$, $r_2(\psi),$ and $r_3(\psi)$ are distinct and not proportional to one another, see Fig.~\ref{fig:jump_rate_rep}(b). (Indeed, the special cases with $r_k(\psi)=0$ are the kernels of ${\cal J}_{1},{\cal J}_2,{\cal J}_3$, which are all distinct.)
In contrast, $J_c (\vartheta )$ and $J_s(\vartheta,\phi )$ both lead to dephasing in the considered basis, they are proportional to the same  operator $J$.

\subsection{SJEDs}
\label{sec:ex_sjed}

From Eq.~\eqref{eq:ex1:rep1} we construct a representation for some fixed $\vartheta,\phi$ as $H,J_1,\dots,J_5$ with $J_4=J_c(\vartheta)$ and  $J_5=J_s(\vartheta,\phi)$.
The corresponding SJEDs are $S_1=\{1,2,3\}$ (of reset type) and  $S_2=\{4,5\}$ (of non-reset type). Their composite actions are
\begin{subequations}\label{eq:ex1:SJED_rep1}
\begin{align}
    {\mathcal{A}}_1(\psi) &= \Tr(\Gamma\psi)\,|0\rangle\!\langle 0|,\quad 
    \\
    {\mathcal{A}}_2(\psi) &= |\lambda|^2 \mathcal{J}(\psi),
\end{align}
\end{subequations}
with
   $ \Gamma/\gamma
    = (1+\cos^2\theta)|1\rangle\!\langle1| 
    +(1+\sin^2\theta)|2\rangle\!\langle 2|
    + \cos\theta\sin\theta(|1\rangle\!\langle 2| + |2\rangle\!\langle 1|)$.
Crucially, these actions ${\cal A}_1,{\cal A}_2$ are independent of the parameters $\vartheta$  and $\phi$. 
Therefore, one may construct a second representation of the same QME by replacing $\vartheta,\phi$ with new values $\tilde{\vartheta}, \tilde{\phi}$.  This representation is denoted $H,\tilde J_1,\dots,\tilde J_5$ with $\tilde J_k=J_k$ for $k=1,2,3$ while $\tilde J_4=J_c(\tilde\vartheta)$ and  $\tilde J_5=J_s(\tilde\vartheta,\tilde\phi)$.

The representations $H, J_1,\dots, J_5$ and $H,\tilde J_1,\dots,\tilde J_5$
obey Theorems~1 and~3 with the trivial permutation, $\pic(\alpha)=\alpha$ for $\alpha=1,2$.  This does not change if the Hamiltonian $H$ is shifted by a real constant for any of the representations. 

However, the representations $H,J_1,\dots,J_5$ and $H,\tilde{J}_1,\dots,\tilde{J}_5$  generically do not respect Theorem~2, but there are special cases:
\begin{align}
	&\tilde{\vartheta}=\vartheta : & \hspace{-5mm} &\tilde{J}_4 =J_4  , & \hspace{-5mm} &\tilde{J}_5 = {\rm e}^{i (\tilde{\phi}-\phi)} J_5  ,	\nonumber\\
	&\tilde{\vartheta}=\vartheta+90^\circ : & \hspace{-5mm} &\tilde{J}_4 = -{\rm e}^{-i\phi} J_5  , & \hspace{-5mm} &\tilde{J}_5 = {\rm e}^{i\tilde{\phi}} J_4 , 
	\nonumber\\
		& \tilde{\vartheta}=\vartheta+180^\circ :& \hspace{-5mm}  &\tilde{J}_4 = - J_4  ,& \hspace{-5mm} &\tilde{J}_5  = - {\rm e}^{i (\tilde{\phi}-\phi)} J_5 ,	\nonumber\\
			&\tilde{\vartheta}=\vartheta+270^\circ :& \hspace{-5mm}  &\tilde{J}_4 = {\rm e}^{-i\phi} J_5 , & \hspace{-5mm} &\tilde{J}_5 = -{\rm e}^{i\tilde{\phi}} J_4 .
		\label{eq:ex1:theorem2}
\end{align}
(We quote angles in degrees to avoid confusion of the permutation $\pi$ with an angle of $\pi$ radians.)
In these cases, Eq.~\eqref{eq:WF_cond} is obeyed (for all $\phi,\tilde\phi$): the permutation $\pi$ is trivial for $\tilde{\vartheta}=\vartheta$ or $\tilde{\vartheta}=\vartheta +180^\circ $.  For $\tilde{\vartheta}=\vartheta +90^\circ $ and $\tilde{\vartheta}=\vartheta +270^\circ $, the permutation swaps $4$ and $5$ (and is trivial otherwise). In these cases, Theorem~2 is valid and the two representations have identical ensembles of labelled quantum trajectories, up to the corresponding permutation of measurement records.

\subsection{Minimal representations for SJEDs}

We now consider another representation that encodes the same QME as $H,J_1,\dots,J_5$.  It is defined such that the composite actions in Eq.~\eqref{eq:ex1:SJED_rep1} have minimal representations.  This is achieved by diagonalising $\Gamma$ and combining proportional jumps $\tilde J_4, \tilde J_5$ (see also Appendix~\ref{app:gauge-sjed}), which fixes the resulting jump operators up to  an overall relabelling (permutation), and multiplication by arbitrary phase factors. We choose this new representation is $H,J'_1,J'_2,J'_3$ with
\begin{equation}\label{eq:ex1:rep2}
    J'_1 =  \sqrt{\gamma}|0\rangle\!\left(-\sin\theta\langle 1|+\cos\theta\langle 2|\right), \quad
    J'_2 = \sqrt{2\gamma}|0\rangle\!\left(\cos\theta\langle 1|+\sin\theta\langle 2|\right), \quad
    J'_3 = \lambda \deph.
\end{equation}
The associated jump rates  are shown in Fig.~\ref{fig:jump_rate_rep}.

The resulting SJEDs are $S'_1=\{1,2\}$ (reset type) and $S'_2=\{3\}$ (non-reset).  These do indeed give rise to the same composite actions as in Eq.~\eqref{eq:ex1:SJED_rep1}, that is, ${\cal A}'_{\alpha}={\cal A}_{\alpha}$ for $\alpha=1,2$, cf.~Fig.~\ref{fig:jump_rate_rep}. Therefore, Theorem~1 is valid for the representations $H,J_1,\dots J_5$ and $H',J'_1,\dots,J'_3$ (with the trivial permutation $\pic$), and their unravelled generators coincide lead to the same quantum trajectory ensembles. 

Furthermore, these two representations are related as ${J}_j = \sum_{k=1}^3 \mathbf{V}_{jk} J_k'$,  where
\begin{equation}
	\mathbf{V} =
	\begin{pmatrix}
		- \sin\theta& \frac{1}{\sqrt{2}} \cos\theta &0\\
	 \cos\theta &\frac{1}{\sqrt{2}}  \sin\theta&0\\
		0&\frac{1}{\sqrt{2}}&0\\
		0&0&\cos \vartheta\\
		0&0&{\rm e}^{i\phi}\sin \vartheta
	\end{pmatrix}.
	\label{eq:ex-V}
\end{equation}
This is a gauge transformation consistent with Eq.~\eqref{eq:V_rep_connection}, as it must be.  One sees that $\mathbf{V}$ is an isometry since its columns are orthonormal; it also decomposes into two blocks $\mathbf{V}^{(1)}$ and $\mathbf{V}^{(2)}$ that represent isometries within the two SJEDs.

The representations $H,J_1,\dots J_5$ and $H',J'_1,\dots,J'_3$  also obey Theorem~3 (with the same permutation $\pic$) and their partially-labelled quantum trajectory ensembles are identical (due to $\pic$ being trivial).  
 However, the representations have a different number of jump operators so they cannot obey Theorem~2 (the ensembles of labelled quantum trajectories cannot be equivalent for the two representations because their measurement records refer to different types of emitted quanta).
 
\subsection{Permutations}

We explained in Sec.~\ref{sec:WF} that Theorem~2 may hold for more than one permutation simultaneously, but that the permutation in Theorem 1 is unique.  We now show how this plays out in our example.  A more complex example is given in Appendix~\ref{app:example}.

Consider first the representations $H,J_1,\dots,J_5$ and $H,\tilde{J}_1,\dots,\tilde{J}_5$. Theorem~2 is valid in the cases outlined in Eq.~\eqref{eq:ex1:theorem2}.  A suitable choice for the permutation $\pi$ was constructed in Sec.~\ref{sec:ex_sjed}: note that it acts as $\pi(k)=k$ for $k=1,2,3$ although it may be non-trivial for $k=4,5$.  This choice of $\pi$ is unique except in special cases such as $\theta=0,180^\circ$, which lead to $J_1=\pm J_3$.  In that case Theorem 2 holds simultaneously for an alternative permutation that swaps $1$ and $3$.
Similarly for $\theta=90^\circ,270^\circ$ then $J_2=\pm J_3$ and an alternative permutation can be constructed.

Now consider the representations $H,J_1,\dots,J_5$ and $H,J'_1,\dots,J'_3$.  As explained above, Theorems 1 and 3 hold here with the unique (trivial) permutation $\pic(\alpha)=\alpha$ for $\alpha=1,2$.  However, one might equivalently have defined the SJEDs for the latter representation as $S_1'=\{3\}$ and $S_2'=\{1,2\}$ in which case $\pic$ would swap indices $1,2$.
In this case, their quantum partially-labelled quantum trajectory ensembles would no longer be identical but remain equivalent. This demonstrates that non-uniqueness associated with labelling of SJEDs does not affect the equivalence of the partially-labelled dynamics (for an example with two reset SJEDs, see Appendix~\ref{app:example}).

We conclude that labelling jump operators or SJEDs for any representation introduces freedom of permuting their types or labels, which is correctly accounted for when considering the gauge equivalence rather than the gauge invariance of the labelled and the partially labelled dynamics. 

\subsection{General representations}

Finally, we observe that the representation in Eq.~\eqref{eq:ex1:rep2} is actually a minimal representation of the QME. Therefore, a general representation of this QME can be obtained by the transformation Eq.~\eqref{eq:QME_transformations}, which shifts and mixes the jumps, together with an appropriate change in the Hamiltonian.  The transformation in Eq.~\eqref{eq:ex-V} features no shifts, and it also has a block structure, as encoded by Eq.~\eqref{eq:V_rep_connection}: this ensures that $H,J_1,\dots,J_5$ and $H,J_1',\dots,J_3'$ lead to identical quantum trajectories, and equivalent partially-labelled quantum dynamics.  In order to ensure equivalent (fully)-labelled quantum dynamics, the isometric mixing of jump operators must reduce to simple permutation and multiplication by phases as in Eq.~\eqref{eq:ex1:theorem2}.

\section{Proof of Theorem~1}\label{sec:proof_1}

In this section, we give the proof of Theorem~1, which was presented in Sec.~\ref{sec:theorem_statement}. \\

\subsection{Conditions for equality of unravelled generators}\label{sec:preliminaries}

Consider two unravelled generators $\mathcal{W}$ and $\tilde{\mathcal{W}}$ whose Hamiltonian and jump operators are $H, J_1, \dots, J_d$ 
and 
$\tilde H, \tilde J_1, \dots, \tilde J_{\tilde d}$.
We derive conditions under which $\tilde{\cal W}={\cal W}$, or equivalently
\beq
\tilde{\mathcal{W}} f(\psi) -{\mathcal{W}} f(\psi)=0
\label{eq:WWf}
\eeq for all functions $f$,
where [cf.~Eq.~\eqref{eq:W}]
\begin{equation}\label{eq:Wt}
    \tilde{\mathcal{W}}f(\psi) = \tilde{\mathcal{B}}(\psi)\cdot\nabla f(\psi)
    +\sum_{j=1}^{\tilde{d}}\int d\psi' \tilde w_j(\psi,\psi')\left[f(\psi')-f(\psi)\right],
\end{equation}
with
$\tilde{\mathcal{B}}(\psi)$ and $\tilde w_j(\psi',\psi)$ defined as in Eqs.~\eqref{eq:B} and~\eqref{eq:w_k} but with $H, J_1, \dots, J_d$ replaced with $\tilde H, \tilde J_1, \dots, \tilde J_{\tilde d}$. 
Then Eq.~\eqref{eq:WWf} is equivalent to
\begin{multline}\label{eq:W_dif}
    0 
   = \left[\tilde{\mathcal{B}}(\psi)-\mathcal{B}(\psi)\right]\cdot \nabla f(\psi) \\ + \int d\psi' \left[  \sum_{j=1}^{\tilde{d}}\tilde{w}_j(\psi,\psi')-\sum_{k=1}^d w_k(\psi,\psi')\right]\left[f(\psi')-f(\psi)\right].
\end{multline}

We will now separate this condition into two conditions for the drift term (proportional to $\nabla f$) and the jump terms (the remaining terms).
To this end, we consider the function
\begin{equation}\label{eq:f}
    f_{\varphi,\epsilon}(\psi) = \exp\left[\frac{\Tr(\varphi\psi)-1}{\epsilon}\right],
\end{equation}
where $\varphi$ is a pure state and $\epsilon>0$.
Taking the gradient with respect to $\psi$, we have
\begin{equation}
    \left[\nabla f_{\varphi,\epsilon}(\psi)\right]_{ab} = \frac{1}{\epsilon}\varphi_{ab}f_{\varphi,\epsilon}(\psi),
\end{equation}
where $a,b$ indicate the relevant matrix elements.
Putting this $f_{\varphi,\epsilon}$ into
\eqref{eq:W_dif}  and multiplying by $\epsilon$, we obtain
\begin{multline}\label{eq:W_dif_f}
    0 = \sum_{ab} \left[\tilde{\mathcal{B}}(\psi)-\mathcal{B}(\psi)\right]_{ab}\varphi_{ab}f_{\varphi,\epsilon}(\psi) \\
    + \epsilon \!\int\! d\psi' \!\!\left[ \! \sum_{j=1}^{\tilde{d}}\tilde{w}_j(\psi,\psi')\!-\!\sum_{k=1}^d w_k(\psi,\psi')\right]\!\!\left[f_{\varphi,\epsilon}(\psi')\!-\!f_{\varphi,\epsilon}(\psi)\right],
\end{multline}
which must hold for all $\epsilon$ and all pure states $\psi$. 
Using that $0<f_{\varphi,\epsilon}\leq 1$, we take $\epsilon\rightarrow 0$ and evaluate at $\psi=\varphi$
to obtain 
    \begin{equation}\label{eq:B0}
        \sum_{ab} \left[\tilde{\mathcal{B}}(\varphi)-\mathcal{B}(\varphi)\right]_{ab}\varphi_{ab} = 0\,.
    \end{equation}    
    for all pure states $\varphi$ (and matrix elements $ab$).  Hence, the drifts are equal, that is

    \begin{equation}\label{eq:B1}
      \tilde{\mathcal{B}}(\psi)=   \mathcal{B}(\psi)  \quad \forall \, \psi \; .
    \end{equation}

   From Eqs.~\eqref{eq:W_dif} and~\eqref{eq:B1} it follows that the jump terms must coincide as well,
    \begin{equation}\label{eq:jump_zero}
       \int d\psi' \left[ \sum_{j=1}^{\tilde{d}}\tilde{w}_j(\psi,\psi')\!-\!\sum_{k=1}^d w_k(\psi,\psi') \right]\!\left[f(\psi')\!-\! f(\psi)\right] = 0
    \end{equation}
    for all pure states $\psi$ and functions $f$. Therefore, this is equivalent to the condition
\begin{equation}\label{eq:summed}
 \sum_{j=1}^{\tilde{d}} \tilde{w}_j(\psi,\psi')=	\sum_{k=1}^d w_k(\psi,\psi')  \quad \forall \, \psi,\psi' \; .
\end{equation}

In summary, we have shown the two conditions in Eqs.~\eqref{eq:B1} and~\eqref{eq:summed} are together equivalent to Eq.~\eqref{eq:WWf}, which in turn means that the unravelled generators are equal for the two representations.
We refer to Eqs.~\eqref{eq:B1} and~\eqref{eq:summed}  as the \emph{drift condition} and the \emph{jump condition}, respectively.
We analyse these conditions separately, before combining the results in Sec.~\ref{sec:final-thm1} to prove Theorem~1.

\subsection{Jump condition}\label{sec:proof_jumps}

The jump condition in Eq.~\eqref{eq:summed} can be expressed using Eq.~\eqref{eq:w_k} as
\begin{equation}\label{eq:w_psic}
\sum_{j=1}^{\tilde d} \delta\!\left[\psi'-\tilde{\mathcal{D}}_j(\psi)\right]\tilde{r}_j(\psi)=	\sum_{k=1}^d \delta\!\left[\psi'-\mathcal{D}_k(\psi)\right]r_k(\psi)
\end{equation}
with rates $r_k(\psi)$ and destinations ${\cal D}_k(\psi)$ as defined in Eqs.~\eqref{eq:r_k} and~\eqref{eq:D_k}. We analogously define  the rates $\tilde{r}_j(\psi)$  and the destinations $\tilde{\mathcal{D}}_j(\psi)$ for the jump operators $\tilde J_j$.

Since both sides of Eq.~\eqref{eq:w_psic} consists of sums of delta functions, equality requires that the sets of destinations are equal on both sides.  Physically, this means that that for two representations to have the same unravelled quantum dynamics, their jump operators must lead to the same destinations.  We formalise this idea using SJEDs and their jump action operators: it turns out that Eq.~\eqref{eq:theorem_cond_J} is a necessary and sufficient condition for Eq.~\eqref{eq:w_psic} to hold.  We show this by first analysing Eq.~\eqref{eq:w_psic} for pure states in a particular set $\cset$, and then using linearity of ${\cal A}_\alpha$ to extend the analysis to all pure states.

\subsubsection{A set of states where every SJED has a distinct destination}
\label{sec:different-dest}

To define the relevant set $\cset$, we require that for any pure state $\psi\in \cset$, the SJEDs all have different destinations (one representation at a time). This requires the following three properties to be fulfilled:
\begin{itemize}
	\item[(i)] the destinations $\mathcal{D}_k(\psi)\neq0$ for all $k$, and similarly $\tilde{\mathcal{D}}_j(\psi)\neq0$ for all $j$;
	
		\item[(ii)] the destinations $\mathcal{D}_k(\psi)\neq \mathcal{D}_{k'}(\psi)$ for all $k\in S_\alpha$ and $k'\in S_{\alpha'}$ with any $\alpha\neq\alpha'$; 
		
			\item[(iii)] the destinations $\tilde{\mathcal{D}}_j(\psi)\neq \tilde{\mathcal{D}}_{j'}(\psi)$ for all $j\in \tilde S_\beta$ $j'\in \tilde S_{\beta'}$ with any $\beta\neq\beta'$.
	
\end{itemize}
{Recall we restrict that $J_j\neq0$ for all $j$.}

To construct $\cset$, we begin with a single state $\psi=\psi_0$ such that properties (i)-(iii) hold.
It is guaranteed by the definition of SJEDs that such a $\psi_0$ always exists.  In fact, the definition is minimal for this to be guaranteed [recall Eq.~\eqref{eq:def-cgs}].
In Appendix~\ref{app:C} we describe a systematic approach for finding a suitable $\psi_0$.

Crucially, the properties of $\psi_0$ already ensure that there is a finite neighbourhood around $\psi_0$ in which other pure states $\psi$  still have distinct destinations with respect to all SJEDs, as follows. 
Specifically, let us take
\beq\label{eq:C}
    \cset = \big\{ \psi  :\,\, 
    \Tr(\psi_0 \psi)  > 1\!-\!\delta^2 \big\},
\eeq
which is the intersection of the set of pure density matrices with the ball in the space of linear operators centred at $\psi_0$ with a radius $\delta>0$ in the trace distance.
We now explain that it is always possible to take $\delta>0$ small enough that the properties (i)-(iii) hold true for all $\psi \in \cset$ in Eq.~\eqref{eq:C}. 

\begin{figure}
    \centering
	\includegraphics[width=0.9\linewidth]{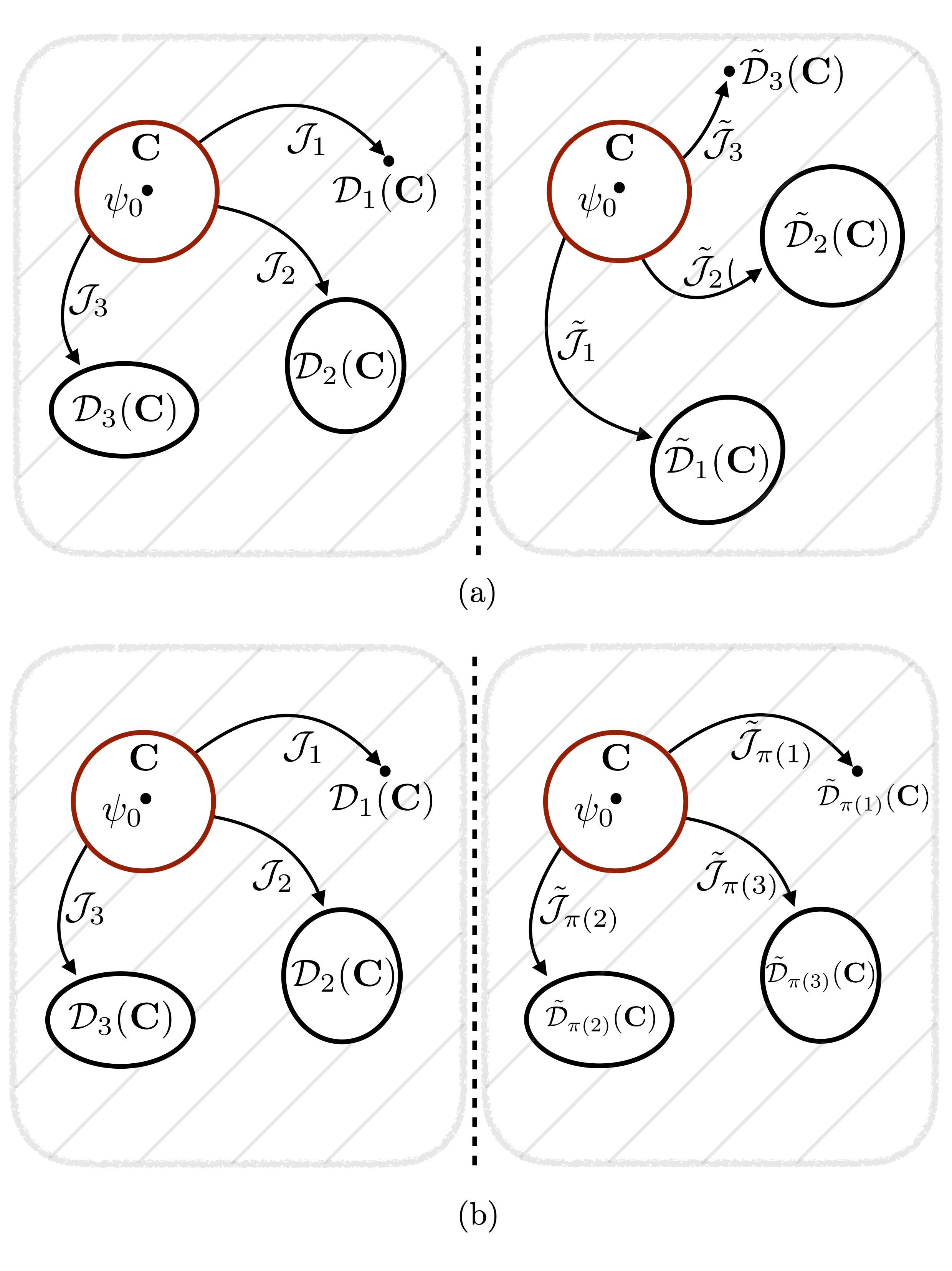}
	\caption{\textbf{(a)} Generic action of jump operators on pure density matrices (gray, shaded) from the set $\cset$ (red).
			The images of the action of all jump operators are disjoint.  Note that $J_1$ is a reset jump operator as it always has the same destination. \textbf{(b)}  For systems compatible with Theorem~2, the images of the jump operators must coincide, up to a permutation (and multiplication by a phase, not indicated as does not contribute to their actions).}
	\label{fig:sketch}
	\phantomlabel{(a)}{fig:sketch_a}
	\phantomlabel{(b)}{fig:sketch_b}
\end{figure}

To this end, write ${\cal D}_k(\cset) = \{ \mathcal{D}_k(\psi) : \psi\in\cset\}$ for the set of destinations from $\cset$ facilitated by jump operator $J_k$, and similarly $\tilde{\cal D}_j(\cset) $ for the analogous set facilitated by jump operator $\tilde J_j$, see Fig.~\ref{fig:sketch}.
To show that (i) holds throughout $\cset$, note that the rates $r_k(\psi_0)$, $\tilde{r}_j(\psi_0)$ are non-zero for all $j$ and $k$: hence by linearity [cf.~Eq.~\eqref{eq:r_k}] there exist sufficiently small $\delta>0$ such that (i) follows for all $\psi\in \cset$ [cf.~Eq.~\eqref{eq:D_k}].  Similarly, (ii) holds for $\psi=\psi_0$ which means that the destinations ${\cal D}_k(\psi_0)$ are different for each SJED. Furthermore, the destinations are continuous functions of $\psi$ provided that their rates are non-zero, which is already guaranteed by (i). Thus there exists  $\delta>0$ [as chosen for (i) or smaller] such that ${\cal D}_k(\cset)$ is disjoint from ${\cal D}_{k'}(\cset)$ if $k\in S_\alpha$ and $k'\in S_{\alpha'}$ with $\alpha\neq\alpha'$, see Fig.~\ref{fig:sketch_a}.  Then indeed the property (ii) is true for all $\psi\in \cset$.  The argument for property (iii) is exactly analogous.

\subsubsection{Singleton SJEDs}\label{sec:singleton}

From the jump condition expressed as Eq.~\eqref{eq:w_psic}, we can expect that the action of SJEDs must lead to the same sets of destinations for both representations.
In this section, we assume that all SJEDs are singletons (i.e., each SJED contains a single jump). Then, in the setting of Fig.~\ref{fig:sketch_a}, this requires that the destination sets $\mathcal {D}_1(\cset),\dots,{\mathcal{D}}_{d}(\cset)$ should be in one-to-one correspondence with $\tilde{\mathcal{D}}_{1}(\cset),\dots,\tilde{\mathcal{D}}_{\tilde{d}}(\cset)$.  Formalising this observation, we derive conditions on the jump operators such that  Eq.~\eqref{eq:w_psic} is satisfied [see Eq.~\eqref{eq:matrix-perm} below].  This is helpful to build physical intuition before  discussing the case of general SJEDs in Sec.~\ref{sec:jump-arb-SJED}.

We start by considering the jump condition in Eq.~\eqref{eq:w_psic} for some $\psi\in\cset$.
From the properties of $\cset$, all the factors (the rates) are strictly positive [see property (i)], and the Dirac delta functions on LHS and RHS are located at different points [the destinations, see properties (ii) and (iii)].  Then the equality requires that the two sums contain the same number of terms, that is 
\begin{equation}\label{eq:d}
\tilde{d}=d.  
\end{equation}
Moreover, the locations of the delta functions must be the same on LHS and RHS, which means that
\begin{equation}\label{eq:D_perm}
     \tilde{\mathcal{D}}_k(\psi) = \mathcal{D}_{\pi_\psi(k)}(\psi)  \quad \forall\, k=1,\dots,d\quad\forall \,\psi\in\cset,
\end{equation}
where $\pi_\psi$ is a permutation of $\{1,\dots,d\}$, dependent in general on $\psi$. It also follows that the rates coincide up to the same permutation,
\begin{equation}\label{eq:r_perm}
	\tilde{r}_k(\psi) = r_{\pi_\psi(k)}(\psi)   \quad \forall\, k=1,\dots,d\quad \forall \,\psi\in\cset.
\end{equation} 
In terms of Fig.~\ref{fig:sketch}(a), this means that the union of $\mathcal {D}_1(\cset),\dots,{\mathcal{D}}_{d}(\cset)$ coincides with the union of $\tilde{\mathcal{D}}_{1}(\cset),\dots,\tilde{\mathcal{D}}_{\tilde{d}}(\cset)$ and the sets of rates coincide as well.

We now prove that the permutation $\pi_\psi$ is in fact  independent of $\psi\in\cset$, that is,
\begin{equation}\label{eq:perm}
	\pi_\psi=\pi ,
\end{equation}
  which means that  the destination sets for the two representations, $\mathcal {D}_1(\cset),\dots,{\mathcal{D}}_{d}(\cset)$ and $\tilde{\mathcal{D}}_{1}(\cset),\dots,\tilde{\mathcal{D}}_{\tilde{d}}(\cset)$,  are themselves related by the permutation $\pi$ [see Fig.~\ref{fig:sketch_b}].  To show this,
fix $k$ and suppose that $\pi_{\psi_1}(k)\neq\pi_{\psi_2}(k)$ for some $\psi_1,\psi_2\in\cset$.
Then Eq.~\eqref{eq:D_perm} means that $\tilde{\mathcal{D}}_k(\psi_1), \tilde{\mathcal{D}}_k(\psi_2)\in \tilde{\mathcal{D}}_k(\cset)$ but  $\tilde{\mathcal{D}}_k(\psi_1)\in \mathcal{D}_{\pi_{\psi_1}(k)}(\cset)$ and $\tilde{\mathcal{D}}_k(\psi_2)\in {\mathcal{D}}_{\pi_{\psi_2}(k)}(\cset)$.  Consider a continuous path $\bm{S}$ between $\psi_1$ and $\psi_2$ that lies within $\cset$.  Parameterising this path by $s\in[0,1]$, we have for every $\psi_s\in \bm{S}$ that the destination $\tilde{\mathcal{D}}_k(\psi_s)\in{\mathcal{D}}_{\pi_{\psi_s}(k)}(\cset)$, by Eq.~\eqref{eq:D_perm}.  However,  as the destinations depend continuously on $\psi\in\cset$,  if $\pi_{\psi_1}(k)\neq\pi_{\psi_2}(k)$  then $\tilde{\mathcal{D}}_k(\bm{S})=\{\tilde{\mathcal{D}}_{k}(\psi_s):\psi_s\in \bm{S}\}$ is a path that connects the two disjoint sets $\mathcal{D}_{\pi_{\psi_1}(k)}(\cset)$ and $\mathcal{D}_{\pi_{\psi_1}(k)}(\cset)$, which is not possible since $\tilde{\mathcal{D}}_k(\bm{S})\in\tilde{\mathcal{D}}_k(\cset)$.  Hence $\pi_{\psi_1}(k)=\pi_{\psi_2}(k)$. As this holds for any $k$ and $\psi_1,\psi_2\in\cset$,  $\pi_\psi$ is indeed independent of $\psi$.
The resulting situation is illustrated in Fig.~\ref{fig:sketch_b}.

The next step uses Eqs.~\eqref{eq:D_perm} and~\eqref{eq:r_perm} to establish conditions on jump operators.
Eqs.~\eqref{eq:r_k} and~\eqref{eq:D_k} relate the jump action ${\cal J}_k$ to the rate $r_k$ and destination ${\cal D}_k$: then using 
Eqs.~\eqref{eq:D_perm} and~\eqref{eq:r_perm} together with Eq.~\eqref{eq:perm} we obtain
\begin{equation}\label{eq:perm1}
     \tilde{\mathcal{J}}_k(\psi) = {\mathcal{J}}_{\pi(k)}(\psi), \quad \forall\, k=1,\dots,d\quad \forall\,\psi\in\cset.
\end{equation}
Recalling that $\psi=|\psi\rangle\!\langle\psi|$ and using Eq.~\eqref{eq:superJ}, this implies
\begin{equation}
\tilde{J}_k |\psi\rangle= {\rm e}^{i\phi_{k}^\psi} J_{\pi(k)} |\psi\rangle  \quad \forall\, k=1,\dots,d \quad \forall\,\psi\in\cset,
\label{eq:perm3}
\end{equation}
where the phase $\phi_{k}^\psi \in \mathbb{R}$ may depend in general on both $k$ and $\psi$.

The final part of this proof is 
to show that the phase in Eq.~\eqref{eq:perm3} does not depend on $\psi$, so this condition holds for all $\psi$ (not just $\psi\in\cset$), and hence
\beq
\tilde{J}_k  = J_{\pi(k)} {\rm e}^{i\phi_k},\quad \forall\, k=1,\dots,d
\label{eq:matrix-perm}
\eeq
with $\phi_k\in\mathbb{R}$.
This result is consistent with the condition in Eq.~\eqref{eq:theorem_cond_J}, for the case of singleton SJEDs [cf.~Eq.~\eqref{eq:single-J-phi}]. 
Eq.~\eqref{eq:theorem_cond_J} also implies that Eq.~\eqref{eq:perm1} holds for any $\psi$.

To show Eq.~\eqref{eq:matrix-perm} is implied by Eq.~\eqref{eq:perm3}, we use linearity of the jump operator $J_k$.  Write
a generic state in $\cset$ as 
\begin{equation}
|\psi_c\rangle = \frac{1}{z_c} \left( |\psi_0\rangle + c|\Delta\rangle \right)
\label{equ:psic}
\end{equation}
with  $c\in \mathbb{C}$ and $z_c$ a normalisation constant.  
Recall that the set $\cset$ is an intersection of pure states with a neighbourhood of $\psi_0$, so for any $|\Delta\rangle$ there exists a finite range of $c$, including $c=0$, such that  $\psi_c\in\cset$.
We consider Eq.~\eqref{eq:perm3} for  $\psi=\psi_0$ and $\psi=\psi_c$, and multiply by $z_0$ and $z_c$, respectively, to get
\begin{align}\label{eq:perm6_0}
&\tilde{J}_{k}|\psi_0\rangle ={\rm e}^{i\phi_0} J_{\pi(k)} |\psi_0\rangle,  \\
&\tilde{J}_{k}|\psi_0\rangle + c\tilde{J}_k |\Delta\rangle ={\rm e}^{i\phi_c} \big[J_{\pi(k)} |\psi_0\rangle  + c J_{\pi(k)} |\Delta\rangle\big],
\label{eq:perm6}
\end{align}
where we abbreviated $\phi_{k}^{\psi_0}=\phi_0$ and $\phi_{k}^{\psi_c}=\phi_c$. Taking the scalar product of Eq.~\eqref{eq:perm6} with itself and rearranging, one obtains 
\begin{multline}\label{eq:perm-norm}
    \langle \psi_0 |  \big[ \tilde{J}_k^\dag \tilde{J}_k -{J}_{\pi(k)}^\dag {J}_{\pi(k)} ] |\psi_0 \rangle
	+ c  \,\langle \psi_0 |  \big[ \tilde{J}_k^\dag \tilde{J}_k -{J}_{\pi(k)}^\dag {J}_{\pi(k)} ] |\Delta \rangle \\
	 + c^*  \langle \Delta |   \big[ \tilde{J}_k^\dag \tilde{J}_k -{J}_{\pi(k)}^\dag {J}_{\pi(k)} ]| \psi_0\rangle
	 +	|c|^2 \langle \Delta | \big[ \tilde{J}_k^\dag \tilde{J}_k -{J}_{\pi(k)}^\dag {J}_{\pi(k)} ] |\Delta \rangle=0 
\end{multline}
This holds for all $|\Delta\rangle$ and sufficiently small (complex) $c$, so the coefficients of $1,c,c^*,|c|^2$ need to match. Taking the terms with $1$ and $c$ yields
\begin{align}\label{eq:perm-norm_0}
	\langle \psi_0 | \tilde{J}_k^\dag \tilde{J}_k  |\psi_0 \rangle=	\langle \psi_0 | {J}_{\pi(k)}^\dag {J}_{\pi(k)} |\psi_0\rangle,\\
	\label{eq:perm-prod}
\langle \psi_0 | \tilde{J}_k^\dag  \tilde{J}_k  |\Delta \rangle=\langle \psi_0 |  {J}_{\pi(k)}^\dag {J}_{\pi(k)} |\Delta \rangle.
\end{align}
Next, taking the scalar product of Eq.~\eqref{eq:perm6_0} with~\eqref{eq:perm6}, we obtain
\begin{equation}\label{eq:jdagj}
	\langle \psi_0 | \tilde{J}_k^\dag  \tilde{J}_k  |\psi_0  \rangle+c\langle \psi_0 | \tilde{J}_k^\dag  \tilde{J}_k  |\Delta \rangle
	={\rm e}^{i(\phi_c-\phi_0)}\left[ \langle \psi_0 | {J}_{\pi(k)}^\dag {J}_{\pi(k)} |\psi_0\rangle+c\langle \psi_0 |  {J}_{\pi(k)}^\dag {J}_{\pi(k)} |\Delta \rangle\right],
\end{equation}
Using Eqs.~\eqref{eq:perm-norm_0} and~\eqref{eq:perm-prod}, and that the LHS is non-zero for small enough $c$ (because $\psi_0\in\cset$),  we arrive at
\begin{equation}\label{eq:perm-phase}
	{\rm e}^{i\phi_0}={\rm e}^{i\phi_c}.
\end{equation}
Finally, putting this back in Eq.~\eqref{eq:perm6} and subtracting Eq.~\eqref{eq:perm6_0} shows that
\beq\label{eq:matrix-perm0}
\tilde{J}_k  |\Delta\rangle = {\rm e}^{i\phi_0} J_{\pi(k)} |\Delta\rangle. 
\eeq
Since $|\Delta\rangle$ can be chosen arbitrarily,  Eq.~\eqref{eq:matrix-perm} follows.

\subsubsection{General SJEDs}
\label{sec:jump-arb-SJED}

We now return to general conditions that are necessary and sufficient for Eq.~\eqref{eq:summed} to be valid.  These are needed for the proof of Theorem~1.  The reasoning is analogous to Sec.~\ref{sec:singleton}.

 Since jumps from the same SJED have the same destination, the jump condition in Eq.~\eqref{eq:w_psic} can be expressed as [cf.~Eqs.~\eqref{eq:SJED_jump} and~\eqref{equ:wA-lemma}]
 \begin{equation}\label{eq:w_psic_A}
 	\sum_{\beta=1}^{\dct} \delta\!\left\{\psi'-\frac{\tilde{\mathcal{A}}_\beta(\psi)}{\Tr[\tilde{\mathcal{A}}_\beta(\psi)]}\right\}\Tr[\tilde{\mathcal{A}}_\beta(\psi)]
 = \sum_{\alpha=1}^{\dc} \delta\!\left\{\psi'-\frac{\mathcal{A}_\alpha(\psi)}{\Tr[\mathcal{A}_\alpha(\psi)]}\right\}\Tr[\mathcal{A}_\alpha(\psi)] 	.
 \end{equation} 
 For $\psi\in\mathbf{C}$, all the Dirac delta functions on LHS have distinct support and non-zero coefficients; the same holds on RHS.  Therefore, we must have [cf.~Eq.~\eqref{eq:d}]
 \begin{equation}\label{eq:\dc}
 \dct=\dc
 \end{equation}
 and following the same reasoning as for the singleton case, we arrive at [cf.~Eq.~\eqref{eq:perm1}]
 \beq
 \tilde{\cal A}_\alpha(\psi) = {\cal A}_{\pic(\alpha)}(\psi) ,\quad\forall \alpha=1,...,\dc\quad \forall \psi\in\cset \,,
 \label{eq:AApi}
 \eeq  
where $\pic$ is a permutation of $\{1,2,\dots,\dc\}$ independent of $\psi\in\cset$.

To prove Theorem~1, Eq.~\eqref{eq:AApi} must be extended to all pure states $\psi$, so that the corresponding super-operators are equal,
\begin{equation}\label{eq:AApi-op}
 \tilde {\cal A}_\alpha = {\cal A}_{\pic(\alpha)} ,\quad\forall \alpha=1,...,\dc,
\end{equation} 
this is Eq.~\eqref{eq:theorem_cond_J}, which appears in Theorem~1.  
We show next that Eq.~\eqref{eq:AApi-op} is indeed implied by Eq.~\eqref{eq:w_psic_A}, and as it is easily verified that the converse also holds, the condition in Eq.~\eqref{eq:AApi-op} is equivalent to the jump condition. To make this last extension we consider separately the two types of SJED. 

{From Eq.~\eqref{eq:AApi} we see that if SJED $\tilde{S}_\alpha$ is of reset type, then $S_{\pic(\alpha)}$ must also be of reset type (and likewise for SJEDs of non-reset type), as the dimension of the image of $\mathcal{A}_\alpha(\psi)$ is $1$ for reset SJEDs and $>1$ for non-reset SJEDs. Hence, if $\alpha$ labels a reset-type SJED, then for $j \in \tilde{S}_\alpha$, we have $\tilde{J}_j = \sqrt{\tilde\gamma_j}|\tilde\rs_\alpha\rangle\!\langle \tilde\xi_j|$ [cf.~Eq.~\eqref{eq:reset}] and from Eq.~\eqref{eq:AApi}
\beq
|\tilde\rs_\alpha\rangle\!\langle\tilde\rs_\alpha|= |\rs_{\pic(\alpha)}\rangle\!\langle\rs_{\pic(\alpha)}|.
 \label{eq:reset-dest-pi}
 \eeq
 Eq.~\eqref{eq:AApi} also implies that
\beq
\Tr(\tilde\Gamma_\alpha  \psi)=\Tr[\Gamma_\pa  \psi].
\label{eq:perm-xi}
\eeq
 Eqs.~\eqref{eq:reset-dest-pi} and~\eqref{eq:perm-xi} hold for all $\psi\in\cset$. 
Parameterising $\psi=|\psi\rangle\!\langle\psi|$ as in Eq.~\eqref{equ:psic}, we recall $\psi_0\in\cset$ and hence obtain from Eq.~\eqref{eq:perm-xi} that 
\begin{equation}
 c \, \langle \psi_0 |\big[\tilde\Gamma_\alpha 
- \Gamma_\pa\big]| \Delta \rangle 
+ c^* \langle\Delta |\big[\tilde\Gamma_\alpha 
- \Gamma_\pa\big] |  \psi_0 \rangle +|c|^2   \langle \Delta | \big[\tilde\Gamma_\alpha 
- \Gamma_\pa\big] | \Delta \rangle =0.
\end{equation}
Analogously to Eq.~\eqref{eq:perm-norm},
since this holds for complex $c$ in a finite neighbourhood of $c=0$ the coefficients for $c,c^*,|c|^2$ need to match.  In particular, from the term with $|c|^2$, by observing that $| \Delta \rangle $ is arbitrary and using Hermiticity of $\tilde\Gamma_\alpha 
- \Gamma_\pa$, we obtain 
\beq\label{eq:Gamma_perm}
\tilde\Gamma_\alpha = \Gamma_\pa,
\eeq
which together with Eq.~\eqref{eq:reset-dest-pi} implies the result in Eq.~\eqref{eq:AApi-op} for reset SJEDs.

We now turn to the non-reset case. If $\alpha$ labels a non-reset SJED, then for $j\in \tilde{S}_{\alpha}$, we have $\tilde{J}_j = \tilde\lambda_j \tilde J^{(\alpha)}$ [recall Eq.~\eqref{eq:non_reset}] and the corresponding composite action is
$
\tilde{\cal A}_\alpha(\psi) = | \tilde\lambda^{(\alpha)}|^2 \tilde{\mathcal{J}}^{(\alpha)} (\psi)
$
with 
$ \tilde \lambda^{(\alpha)} =  \sqrt{ \sum_{j \in \tilde S_{\alpha}} |\tilde\lambda_j|^2 } \; $
 [cf.~Eq.~\eqref{eq:A-lambda}].
Then Eq.~\eqref{eq:AApi} for $\psi\in\cset$ implies
\beq
|\tilde{\lambda}^{(\alpha)}|^2 \tilde{\mathcal{J}}^{(\alpha)} (\psi)  
=  |\lambda^{[\pic(\alpha)]}|^2 \mathcal{J}^{[\pic(\alpha)]} (\psi) ,
\eeq
Thus,
\beq
\tilde{\lambda}^{(\alpha)} \tilde J^{(\alpha)} |\psi\rangle = {\rm e}^{i\phi_{\alpha\psi}}  \lambda^{[\pic(\alpha)]} J^{[\pic(\alpha)]} |\psi\rangle  ,
\eeq
with $\phi_{\alpha}^{\psi} \in \mathbb{R}$ in general dependent on $\alpha$ and $\psi$ [cf.~Eqs.~\eqref{eq:perm1} and~\eqref{eq:perm3}].
 Repeating the analysis of Eqs.~(\ref{eq:perm6_0}-\ref{eq:matrix-perm0}) shows that $\phi_{\alpha}^{\psi}$ does not depend on $\psi$ and establishes the operator equation
\beq
\tilde{\lambda}^{(\alpha)} \tilde J^{(\alpha)}
= {\rm e}^{i\phi_{\alpha}}  \lambda^{[\pic(\alpha)]} J^{[\pic(\alpha)]} 
\label{eq:prop_cond}
\eeq
where $\phi_\alpha\in\mathbb{R}$  [cf.~Eq.~\eqref{eq:matrix-perm}].  This implies Eq.~\eqref{eq:AApi-op} for non-reset SJEDs. Therefore, this result is valid for all SJEDs, as promised.

\subsection{Drift term}\label{sec:drift_proof}

We now consider the drift condition in Eq.~\eqref{eq:B1}. We first show that this condition implies
\begin{equation}\label{eq:Heff_cond}
	\tilde{H}_\text{eff}  =  H_\text{eff} + z\mathbb{1}, \quad z\in\mathbb{C}.
\end{equation}
Then using the jump condition as formulated in Eq.~\eqref{eq:AApi-op}, we show that 
\begin{equation}\label{eq:H_cond}
	\tilde{H}=H+r\mathbb{1}, \quad r\in\mathbb{R}.
\end{equation}
This directly coincides with the condition in Eq.~\eqref{eq:theorem_cond_H} of Theorem~1.

To show first Eq.~\eqref{eq:Heff_cond}, we use Eq.~\eqref{eq:B1} with the definitions of ${\cal B}(\psi)$ and $\tilde{\cal B}(\psi)$ [cf.~Eq.~\eqref{eq:B}] to obtain that
\newcommand{\dif}{V}
\begin{equation}
	\dif\psi-\psi \dif^\dag - \psi \Tr(\dif\psi -\psi \dif^\dag) = 0
    \label{eq:dif}
\end{equation}
for all $\psi$, where $\dif=\tilde{H}_\text{eff}-H_\text{eff}$.
Let $|a\rangle\neq |b\rangle$  be two elements of an orthonormal basis for the system. Take $\psi=|a\rangle\!\langle a|$, 
and multiply Eq.~\eqref{eq:dif} from the left by $\langle b|$ and from the right by $|a\rangle$.  As $\langle a|b\rangle=0$,  Eq.~\eqref{eq:dif} then yields (from its first term) $\langle b|\dif|a\rangle=0$. As this holds for any pair of elements of the basis, $\dif$ is diagonal.  
Furthermore, since this holds for any orthogonal basis, we have
\begin{equation}
V=z\mathbb{1},
\end{equation}
for some complex constant $z$, which result  is equivalent to Eq.~\eqref{eq:Heff_cond}.

To arrive at Eq.~\eqref{eq:H_cond}, we sum Eq.~\eqref{eq:AApi-op} over $\alpha=1,\dots \dc$ to obtain [cf.~Eq.~\eqref{eq:superA}]
\begin{equation} \label{equ:JJ-tot-0}
	 \sum_{j=1}^{\tilde{d}}\tilde{\mathcal{J}}_j (\psi)= \sum_{k=1}^d \mathcal{J}_k (\psi)
\end{equation}
for any $\psi$. Then, considering the trace of Eq.~\eqref{equ:JJ-tot-0} and the fact that $\psi$ is arbitrary, we also have
\begin{equation}
    	 \sum_{j=1}^{\tilde{d}}\tilde{J}_j^\dag \tilde{J}_j = \sum_{k=1}^d J_k^\dag J_k \,.
    \label{equ:JJ-tot}
\end{equation} 
Recalling the definitions of  $\tilde{H}_{\rm eff}$ and $H_{\rm eff}$ [cf.~Eq.~\eqref{eq:Heff}], one recognises the LHS and the RHS of Eq.~\eqref{equ:JJ-tot} as their anti-Hermitian parts. Thus, Eq.~\eqref{eq:Heff_cond} holds for their Hermitian parts only, that is, $\tilde{H}$ and $H$, respectively,  but then the constant $z$ must be real, which yields Eq.~\eqref{eq:H_cond}.

\subsection{Final result}
\label{sec:final-thm1}

We collect the results obtained so far in order to prove Theorem~1.  Sec.~\ref{sec:preliminaries} showed that ${\cal W}=\tilde{\cal W} \Leftrightarrow (\ref{eq:B1},\ref{eq:jump_zero})$.
Sec.~\ref{sec:proof_jumps} established that $\eqref{eq:jump_zero}  \Leftrightarrow \eqref{eq:AApi-op}$.
Sec.~\ref{sec:drift_proof} established that $(\ref{eq:B1},\ref{eq:jump_zero}) \Rightarrow \eqref{eq:H_cond}$.
Using the definitions in Eqs.~\eqref{eq:B} and~\eqref{eq:w_k},
one straightforwardly checks that $(\ref{eq:AApi-op},\ref{eq:H_cond})\Rightarrow  (\ref{eq:B1},\ref{eq:jump_zero})$.  
Hence we have shown that
\beq
\tilde{\cal W}={\cal W} \; \Leftrightarrow \; (\ref{eq:jump_zero},\ref{eq:B1})  \; \Leftrightarrow \; (\ref{eq:AApi-op},\ref{eq:H_cond})
\eeq
This proves Theorem~1 because Eqs.~\eqref{eq:H_cond} and~\eqref{eq:AApi-op} exactly match Eqs.~\eqref{eq:theorem_cond_H} and~\eqref{eq:theorem_cond_J}.

\section{Proofs of Theorems 2 and 3}
\label{sec:labelled_gen_proof}

This section outlines the proofs of Theorems 2 and 3. They have analogous structure to the proof of Theorem~1 in Sec.~\ref{sec:proof_1}.

\subsection{Proof of Theorem~2}\label{sec:WF_proof}

We derive the conditions under which Eq.~\eqref{eq:PiWf} holds. The gauge equivalence can be expressed as
\beq
\Pi \tilde{\mathcal{W}}_F \Pi^\dagger f(\psi,\bm q) -{\mathcal{W}_F} f(\psi,\bm q)=0.
\label{eq:WFWFf}
\eeq
Taking $f(\psi,\bm q) = g(\psi)$, we obtain Eq.~\eqref{eq:WWf}, i.e., the gauge invariance for the unlabelled dynamics as  $\Pi$ and $\Pi^\dagger $ acts on $\bm q$ only, which condition is equivalent to the drift condition in Eq.~\eqref{eq:B1} and the jump condition in Eq.~\eqref{eq:summed}. From Sec.~\ref{sec:proof_1}, we have already have that the condition in Eq.~\eqref{eq:theorem_rep} in Theorem 1 is necessary for this.  However, the conditions for Theorem~2 are stronger, and we prove below that they are indeed necessary.  

Expanding and re-arranging Eq.~\eqref{eq:WFWFf} gives
\begin{multline}\label{eq:Wf_dif}
    0 = \left[\tilde{\mathcal{B}}(\psi)-{\cal B}(\psi)\right]\cdot\nabla f(\psi,\bm{q}) 
    +  \sum_k\int d\psi' \Big\{\tilde{w}_k(\psi,\psi')\big[f(\psi',\bm{q}+\bm{e}_{\pi(k)})-f(\psi,\bm{q})\big] \\
    - w_k(\psi,\psi')\big[f(\psi',\bm{q}+\bm{e}_{k})-f(\psi,\bm{q})\big]\Big\}.
\end{multline}
Let the function
\begin{equation}
    f(\psi,\bm q) = g(\psi)(\bm{q})_\ui,
\end{equation}
so that 
$f(\psi',\bm{q}+\bm{e}_k) - f(\psi,\bm{q})
    = \left[g(\psi')-g(\psi)\right](\bm q)_\ui +g(\psi')\,\delta_{\ui k}.$
Considering this function in Eq.~\eqref{eq:Wf_dif}, we obtain
\begin{multline}\label{eq:func_into_Wf_dif}
    0 = \left[\tilde{\mathcal{B}}(\psi)-{\cal B}(\psi)\right]\cdot\nabla g(\psi)(\bm{q})_\ui \\
    + \int d\psi' \Bigg\{\sum_k\Big[\tilde{w}_k(\psi,\psi')-w_k(\psi,\psi')
    \Big]\left[g(\psi')-g(\psi)\right](\bm q)_\ui\\
    +\Big[\tilde{w}_{\pi^{-1}(\ui)}(\psi,\psi')-w_\ui(\psi,\psi')\Big]g(\psi')\Bigg\} \,.
\end{multline}
Combining this with Eq.~\eqref{eq:W_dif} multiplied by $(\bm q)_\ui$, the first two lines cancel, and we find
\begin{equation}
    0 = \int d\psi' \Big[\tilde{w}_{\pi^{-1}(\ui)}(\psi,\psi')-w_\ui(\psi,\psi')\Big]g(\psi').
\end{equation}
Since this must hold for all functions $g(\psi)$ and for all $\ui$, taking $j=\pi(k)$, we arrive at an additional necessary condition beyond the drift and jump conditions in Eqs.~\eqref{eq:B1} and~\eqref{eq:summed}, namely,
\begin{equation}
  \tilde{w}_{k}(\psi,\psi') =    w_{\pi(k)}(\psi,\psi') \quad \forall k=1,\dots,d, \quad\forall \psi,\psi',
    \label{eq:wws}
\end{equation}
which we refer to as the \emph{labelled jump condition}.
We note that the labelled jump condition implies the jump condition, and we conclude that the drift and labelled jump condition are necessary for Eq.~\eqref{eq:PiWf} to hold.

The labelled jump condition condition can also  be expressed as [cf.~Eq.~\eqref{eq:w_psic}]
\begin{equation}\label{eq:w_psic_F}
\delta\!\left[\psi'-\tilde{\mathcal{D}}_k(\psi)\right]\tilde{r}_k(\psi)= \delta\!\left[\psi'-\mathcal{D}_{\pi(k)}(\psi)\right]r_{\pi(k)}(\psi)
\end{equation}
for $k=1,\dots, d$ and all $\psi,\psi'$. Note that the permutation $\pi$ in Eq.~\eqref{eq:wws} is determined by $\Pi^\dagger$ in Eq.~\eqref{eq:WFWFf} [cf.~Eq.~\eqref{eq:Pi}] (and not state dependent).
Recalling the arguments of Sec.~\ref{sec:singleton}
one sees that the labelled jump condition in Eq.~\eqref{eq:w_psic_F} requires both
 Eq.~\eqref{eq:D_perm} and~\eqref{eq:r_perm} to hold [together with Eq.~\eqref{eq:perm}], for all $\psi$. 
 Hence, Eq.~\eqref{eq:perm1} holds for any $\psi$, and we obtain [recall~Eq.~\eqref{eq:single-J-phi}]
\begin{equation}\label{eq:J_perm_labelled}
    \tilde{J}_k = {\rm e}^{i\phi_k}J_{\pi(k)}, \quad \forall k=1,\dots,d
\end{equation}
where $\phi_k \in \mathbb{R}$.

We have thus shown that \eqref{eq:PiWf} $\Rightarrow$ (\ref{eq:theorem_rep},\ref{eq:J_perm_labelled}) which together imply condition in Eq.~\eqref{eq:WF_cond} of Theorem~2.  
The converse can be verified by direct calculation.  Hence, Theorem~2 is proven.

\subsection{Proof of Theorem~3}\label{sec:WC_proof}

Similar to the previous section, we now derive the conditions under which Eq.~\eqref{eq:PiWc} holds. The gauge equivalence can be expressed as [cf.~Eq.~\eqref{eq:WWf}]
\beq
\Pi \tilde{\mathcal{W}}_C \Pi^\dagger f(\psi,\bm Q) -{\mathcal{W}_C} f(\psi,\bm Q)=0,
\label{eq:WCWCf}
\eeq
and taking $f(\psi,\bm Q) = g(\psi)$ we recover again Eq.~\eqref{eq:WWf}. 

We can expand and re-arrange Eq.~\eqref{eq:WCWCf}, to obtain 
\begin{multline}\label{eq:Wc_dif}
    0 = \left[\tilde{\mathcal{B}}(\psi)-{\cal B}(\psi)\right]\cdot\nabla f(\psi,\bm{Q}) \\
    + \sum_\alpha\int d\psi'  \Big\{\tilde{W}_\alpha(\psi,\psi')\big[f(\psi',\bm{Q}+\bm{E}_{\pic(\alpha)})-f(\psi,\bm{Q})\big] \\
    - W_\alpha(\psi,\psi')\big[f(\psi',\bm{Q}+\bm{E}_{\alpha})-f(\psi,\bm{Q})\big]\Big\}
\end{multline}
[cf.~Eq.~\eqref{eq:Wf_dif}]. Then,
we consider the function 
\begin{equation}
    f(\psi,\bm Q)
    = [g(\psi')-g(\psi)](\bm{Q})_\beta,
\end{equation}
 so that
  $f(\psi',\bm{Q}+\bm{E}_{\alpha})-f(\psi,\bm{Q})
  = [g(\psi')-g(\psi)](\bm Q)_\beta+g(\psi')\,\delta_{\alpha\beta}.$
Putting this into Eq.~\eqref{eq:Wc_dif} gives
\begin{multline}\label{eq:func_into_Wc_dif}
    0 = \left[\tilde{\mathcal{B}}(\psi)-{\cal B}(\psi)\right]\cdot\nabla g(\psi)(\bm Q)_\beta \\
    + \int d\psi' \Bigg\{\sum_\alpha\left[\tilde W_\alpha(\psi,\psi')-W_\alpha(\psi,\psi')\right]\left[g(\psi')-g(\psi)\right](\bm Q)_\beta \\
    +\left[\tilde W_{\pi^{-1}(\beta)}(\psi,\psi')-W_\beta(\psi,\psi')\right] g(\psi')\Bigg\}.
\end{multline}
Combining this with Eq.~\eqref{eq:W_dif}, 
only the last term in Eq.~\eqref{eq:func_into_Wc_dif} survives, and considering $\beta=\pic(\alpha)$ we obtain the \emph{partially-labelled jump condition},
\begin{equation}
	\tilde{W}_\alpha(\psi,\psi') = W_{\pic(\alpha)}(\psi,\psi') \quad \forall \alpha=1,\dots,\dc, \quad\forall \psi,\psi' \,. 
\end{equation}
We note that the partially-labelled jump condition also implies the jump condition, and we conclude that the drift and partially-labelled jump condition are necessary for Eq.~\eqref{eq:PiWc} to hold.

The partially-labelled jump condition is equivalent to [cf.~Eq.~\eqref{eq:w_psic_A}]
\begin{equation}\label{eq:w_psic_C}
\delta\!\left\{\psi'-\frac{\tilde{\mathcal{A}}_\alpha(\psi)}{\Tr[\tilde{\mathcal{A}}_\alpha(\psi)]}\right\}\Tr[\tilde{\mathcal{A}}_\alpha(\psi)]=\delta\!\left\{\psi'-\frac{\mathcal{A}_{\pic(\alpha)}(\psi)}{\Tr[\mathcal{A}_{\pic(\alpha)}(\psi)]}\right\}\Tr[\mathcal{A}_{\pic(\alpha)}(\psi)]. 
\end{equation}
for all $\psi$.  Hence Eq.~\eqref{eq:AApi} is valid for all $\psi$ and thus
\begin{equation}\label{eq:J_perm_labelled2}
	\tilde{\mathcal{A}}_\alpha = \mathcal{A}_{\pic(\alpha)}, \quad \forall \alpha=1,\dots,\dc \,.
\end{equation}

We have shown that \eqref{eq:PiWc} $\Rightarrow$ (\ref{eq:theorem_rep},\ref{eq:J_perm_labelled2}) which together imply the conditions in Eq.~\eqref{eq:Wc_cond} of Theorem~3.    The converse can be verified by direct calculation.  Hence, Theorem~3 is proven.

\section{Discussion and outlook}\label{sec:conclusion}

Theorems 1-3 characterise situations where different representations of a QME lead to the same stochastic dynamics for the conditional state, and for measurement records.  As an immediate application, they clarify remaining experimental freedoms if one aims to produce particular ensembles of quantum trajectories. 

For trajectories of the conditional state, an important role was played by SJEDs: Theorem~1 states that the quantum trajectory ensembles are equal when the two representations have the same set of super-operators for the action of the SJED, that is $\{{\cal A}_\alpha\}=\{\tilde{\cal A}_\alpha\}$.
This condition is non-trivial because the same super-operator ${\cal A}_\alpha$ has many possible decompositions in terms of jump operators.  For non-reset SJEDs these decompositions are simple in that they require different jump operators to be proportional to each other.  However, for reset SJEDs, there are other (gauge) freedoms for the jump operators.

If one considers labelled quantum trajectories (that is, trajectories together with measurement records) stricter conditions are required for equivalent stochastic dynamics.  Theorem~2 states that the only remaining gauge freedom in this case is that jump operators can be permuted between the representations, and mutliplied by arbitrary phase factors.  However, Theorem~3 shows that if one coarse-grains the measurement record in a suitable way, the equality of quantum trajectory ensembles implies the equivalence of partially-labelled quantum trajectories, and vice versa.

These results provide a theoretical framework to investigate new phenomena in stochastic dynamics of open quantum systems. In particular, they are relevant when assessing whether a given phenomenon is robust to choosing different representations of the quantum master operator or continuous measurement schemes, or if its conditions are more restrictive.  
Any thermodynamic description that depends only on properties of quantum trajectories is naturally invariant under the gauge transformations discussed here~\cite{Menczel2020,Manzano2022}, compare~\cite{Almeida2024}.
Similarly, any dynamical parameters encoded by gauge transformations of unravelled dynamics cannot be inferred from quantum trajectories or coarse-grained measurement records, instead requiring full measurement records \cite{Gammelmark2014,Guta2017,Bompais22,Radaelli24}.
Another relevant example is when a model exhibits weak unitary symmetry and the operation of the symmetry generates a new representation of the same QME. This symmetry remains at the level of the unravelled quantum dynamics when the symmetry transformed representation has a generator which is the same as that for the initial representation. This is studied in \cite{Usymm} which exploits directly the theorems given in this paper. The results presented here could also be useful for study of non-unitary symmetries exhibited by open quantum systems \cite{roberts21,lieu_quad20,Lieu_breaking20,Sa2023}.

\begin{acknowledgments}
{This work was supported by the Engineering and Physical Sciences Research Council [grant number EP/T517847/1].}
\end{acknowledgments}

\appendix

\section{SJEDs and their representations}
\label{app:gauge-sjed}

Here, we first prove that SJEDs are either of the reset type as in Eq.~\eqref{eq:reset} or the non-reset type as in Eq.~\eqref{eq:non_reset}. Then, we prove that Eq.~\eqref{eq:theorem_rep} is equivalent to Eq.~\eqref{eq:V_rep_connection}.

\subsection{Types of SJEDs}
\label{app:sjed}

To see that all SJEDs must obey either Eq.~\eqref{eq:reset} or Eq.~\eqref{eq:non_reset} suppose that $J_k,J_{k'}\in S_\alpha$ and choose $|\psi_1\rangle,|\psi_2\rangle$ such that $J_k|\psi_{1,2}\rangle\neq0$.  Then Eq.~\eqref{eq:def-cgs} implies
\begin{align}
    J_k |\psi_1\rangle &= c_1 J_{k'} |\psi_1\rangle, \nonumber  \\
    J_k |\psi_2\rangle &= c_2 J_{k'} |\psi_2\rangle. 
    \label{eq:JJcc}
\end{align}
But  Eq.~\eqref{eq:def-cgs} also implies $J_k( |\psi_1\rangle+|\psi_2\rangle) = c_3 J_{k'} ( |\psi_1\rangle+|\psi_2\rangle) $ which combined with \eqref{eq:JJcc} gives
\beq\label{eq:JccJcc}
J_{k'} ( c_1 |\psi_1\rangle + c_2 |\psi_2\rangle ) = J_{k'} ( c_3 |\psi_1\rangle+ c_3 |\psi_2\rangle).
\eeq
Multiplying, separately, from the left by state vectors orthogonal to $J_{k'}|\psi_1\rangle$ and $J_{k'}|\psi_2\rangle$, this can be satisfied in two ways: either $c_1=c_2=c_3$, or $J_{k'}|\psi_1\rangle$ and $J_{k'}|\psi_2\rangle$ are parallel such that $J_{k'}|\psi_2\rangle= c_4 J_{k'}|\psi_1\rangle$ with $c_3=(c_1+c_2 c_4)/(1+c_4)$.
Fixing $|\psi_1\rangle$, these results must hold for all $|\psi_2\rangle$ with $J_k |\psi_2\rangle \neq 0$.
If $c_1=c_2$ for all $|\psi_2\rangle$, then from \eqref{eq:JJcc} we have that $J_{k}= c_1 J_{k'}$. 
Otherwise $J_{k'}|\psi_1\rangle$ must be always parallel to $J_{k'}|\psi_2\rangle$ (which in turn is parallel to $J_k|\psi_2\rangle$), therefore $J_k,J_{k'}$ both have rank 1, and the SJED is of reset type.
Note from the definition of SJED types that we choose $J_{k'}|\psi_1\rangle$ and $J_k'|\psi_2\rangle$ being parallel to take precedence over $c_1=c_2$, so that SJEDs contain either only reset or non-reset jumps, according to their type.

\subsection{Representations of Individual SJEDs}
\label{app:gauge-indiv}

We summarise the gauge freedoms of the SJED action [cf.~Eq.~\eqref{eq:superAJ}] 
operator ${\cal A}_\alpha$, defined in Eq.~\eqref{eq:superA}.
A minimal representation of this operator is denoted as
\beq  {\cal A}_\alpha(\psi)=\sum_{k\in S_\alpha'} J_k' \psi  J_k'^\dag  , \eeq
while a generic representation is written as
\beq 
{\cal A}_\alpha(\psi)=\sum_{j \in S_\alpha} {J}_j \psi  {J}_j^\dag .
\eeq
We consider the two types of SJED in turn.

For a non-reset SJED as in Eq.~\eqref{eq:non_reset}, a minimal representation has $|S_\alpha'| = 1$ and
\beq J'_k = \lambda^{(\alpha)} J^{(\alpha)}\eeq
with $k\in S_\alpha'$, $\lambda^{(\alpha)}>0$ and $\Tr[ J^{(\alpha)\dagger} J^{(\alpha)}]=1 $.  
A generic representation has
\beq 
{J}_j=\lambda_j J^{(\alpha)}, \qquad j \in S_\alpha
\eeq 
with $\sqrt{\sum_{j \in S_\alpha} |\lambda_j|^2 }= \lambda^{(\alpha)}$.  
For future convenience, we write this formula as 
\beq  {J}_j = \sum_{k \in S_\alpha'} \hat{\mathbf{V}}_{jk}^{(\alpha)} J_k', \label{equ:isom-nonR} \eeq
with
$ \sum_{j\in S_\alpha} |\hat{\mathbf{V}}^{(\alpha)}_{jk}|^2 =1$ 
for $k\in S_\alpha'$.  (Note, the symbol $\hat{\mathbf{V}}^{(\alpha)}_{jk}$ is only defined for $j \in S_\alpha, \,k \in S_\alpha'$ and we have $|S_\alpha'|=1$ so this formalism seems  unnecessarily cumbersome at this point, but it will be useful below.)

For a reset SJED as in Eq.~\eqref{eq:A-gamma}, diagonalising the matrix $\Gamma_\alpha$ gives a minimal representation with $|S_\alpha'| = d_\alpha'={\rm rank}(\Gamma_\alpha)$  and
 jump operators
 $${J}_k' = \sqrt{\gamma_k'} |\rs_\alpha\rangle\!\langle \xi_k'|,\qquad k \in S_\alpha',$$
 where
$|\xi'_{k}\rangle$ is a normalised eigenvector of $\Gamma_\alpha$ with eigenvalue $\gamma_k'>0$. 
A generic representation of ${\cal A}_\alpha$ has
\beq 
{J}_j = \sqrt{\gamma_j} |\rs_\alpha\rangle\!\langle \xi_j| ,\qquad j  \in S_\alpha
\eeq
with $\gamma_j>0$ and state vectors
$|\xi_{j}\rangle$
such that
$ \sum_{j\in S_\alpha} \gamma_j |\xi_{j}\rangle\langle \xi_j|=\Gamma_\alpha $ and $\langle \xi_j|\xi_j\rangle=1$.
The vectors
$|\xi_{j}\rangle$
are not orthogonal in general and can be expressed 
as~\cite{Wolf2012} 
\beq
\sqrt{\gamma_j}|\xi_j\rangle  = \sum_{k\in S_\alpha'} [\hat{\mathbf{V}}_{jk}^{(\alpha)}]^* \sqrt{\gamma_k'} |\xi_{k}'\rangle , \qquad j \in S_\alpha,
\label{equ:isom-res_xi}
\eeq
where $\hat{\mathbf{V}}_{jk}^{(\alpha)}$ describes isometric mixing in the sense that
$ \sum_{j\in S_\alpha} [\hat{\mathbf{V}}^{(\alpha)}_{jk}]^* \hat{\mathbf{V}}^{(\alpha)}_{jk'}=\delta_{kk'}$ for $k,k' \in S'_\alpha$.
(As before, the symbol $\hat{\mathbf{V}}_{jk}^{(\alpha)}$ is only defined for $j\in S_\alpha$ and $k\in S'_\alpha$.)
Therefore,
\beq 
{J}_j = \sum_{k \in S_\alpha'} \hat{\mathbf{V}}_{jk}  J_k', \qquad j   \in S_\alpha .
\label{equ:isom-res}
\eeq
This freedom includes
cases where two or more of $|\xi_j\rangle$ are parallel, in which case one recovers also  freedoms already found in the non-reset case, see Eq.~\eqref{equ:isom-nonR}.  Hence we have shown that generic representations of ${\cal A}_\alpha$ are given by Eqs.~\eqref{equ:isom-nonR} and \eqref{equ:isom-res}.

\subsection{Combined Representations of SJEDs}
\label{app:gauge-comb}

We now consider situations with more than one SJED.
As in Sec.~\ref{sec:gauge-cgs} we suppose that $H', J_1',\dots, J_{d'}'$ is a representation of the QME in which all SJEDs have minimal representations.  There are $\dc$ SJEDs which are $S'_1,\dots,S'_{\dc}$.  Let  ${H},  J_1,\dots,  J_{ d}$ be a generic representation with SJEDs $S_1,\dots,S_{\dc}$.  We will show that conditions Eq.~\eqref{eq:theorem_rep} of Theorem~1 are equivalent to the conditions in Eq.~\eqref{eq:V_rep_connection}.

It can be verified by direct calculation that Eq.~\eqref{eq:V_rep_connection} implies Eq.~\eqref{eq:theorem_rep}.  We now show the converse, that  Eq.~\eqref{eq:theorem_rep} implies \eqref{eq:V_rep_connection}.
Condition~\eqref{equ:H-V-rep} is immediate from Theorem~1.  To establish  the remaining conditions, we construct general jump operators that preserve the set $\{{\cal A}_\alpha\}_{\alpha=1}^{\dc}$, by applying Eqs.~\eqref{equ:isom-nonR} and \eqref{equ:isom-res} with arbitrary isometries.

To do this in a systematic way, we start with an arbitrary partitioning of the indices $1,\dots, d$ into sets ${S}_1,\dots,{S}_{\dc}$.  Choose an arbitrary permutation $\pic$ of $\{1,\dots,\dc\}$.  Then jump operator $ J_j$ can be obtained as
\beq
{J}_j = \sum_{k\in S'_{\pic(\alpha)}}\hat{\mathbf{V}}^{(\alpha)}_{jk} J_k'\quad \text{for} \quad j\in {S}_\alpha,
\label{equ:new-rep-big}
\eeq
where $\hat{\mathbf{V}}^{(\alpha)}$ is similar to the isometry from Eqs.~\eqref{equ:isom-nonR} and~\eqref{equ:isom-res}, 
except that it is now defined for $k\in S'_{\pic(\alpha)}$ [instead of $k\in S'_\alpha$], to allow for permutation of the SJEDs.
The relevant isometric property becomes
\beq
 \sum_{j\in  S_\alpha} \left[\hat{\mathbf{V}}^{(\alpha)}_{jk}\right]^*\hat{\mathbf{V}}^{(\alpha)}_{jk'}=\delta_{kk'}  \quad\text{for}\quad k,k'\in S'_{\pic(\alpha)} \; .
 \label{equ:new-isom}
\eeq

It is useful to embed $\hat{\mathbf{V}}^{(\alpha)}$, 
into a $\tilde{d}\times d$ matrix $\mathbf{V}^{(\alpha)}$, by padding the remaining elements with zeros,
\begin{equation}
	\mathbf{V}_{jk}^{(\alpha)} =  
	\begin{cases} \hat{\mathbf{V}}_{jk}^{(\alpha)}  \quad & \text{if} \quad  j\in  S_\alpha, \, k\in S'_{\pic(\alpha)} \\ 0 & \text{otherwise} \; .
	\end{cases}
	\label{eq:embed-hatV}
\end{equation}
Then [cf.~Eq.~\eqref{equ:new-isom}],
\beq
{J}_j = \sum_{k\in S'_{\pic(\alpha)}}\mathbf{V}^{(\alpha)}_{jk} J_k'\quad \text{for} \quad j\in {S}_\alpha,
\label{equ:new-isom-embed}
\eeq
and summing over SJEDs, we obtain 
\beq
 J_j = \sum_{k=1}^{d} \mathbf{V}_{jk} J_k'
\label{equ:V-rep-app}
\eeq 
with $\mathbf{V} = \sum_{\alpha=1}^{\dc} {\mathbf{V}}^{(\alpha)}$.  Using also Eq.~\eqref{equ:new-isom}, can one can show that $\mathbf{V}$ is isometric (that is $\mathbf{V}^\dag  \mathbf{V}=\mathbb{1}$).

This systematic construction yielded the conditions in Eqs.~\eqref{equ:V-rep-app},~\eqref{eq:embed-hatV}, and~\eqref{equ:new-isom}, which match Eqs.~\eqref{equ:V-V-rep},~\eqref{equ:pad-V-rep}, and~\eqref{eq:V-big-isom}.  Overall, we have established that Eq.~\eqref{eq:theorem_rep} is equivalent to Eq.~\eqref{eq:V_rep_connection}.

\section{Additional Example}\label{app:example}

We discuss an example in which the permutation appearing in Theorems~1 and~3 is non-trivial.
As in Sec.~\ref{sec:example}, we consider a 3-level system with the basis $|0\rangle$, $|1\rangle$, $|2\rangle$ and with an arbitrary Hamiltonian $H$. Let the jump operators be
\begin{align}
	J_1 &= \sqrt{\gamma_1}|\chi_1\rangle\!\langle1|, \quad
	J_2 = \sqrt{\gamma_2}|\chi_1\rangle\!\langle1|,\quad J_3 = \sqrt{\gamma_3}|\chi_1\rangle\!\langle2|, \nonumber \\
	J_4 &= \sqrt{\gamma_1+\gamma_2}|\chi_2\rangle\!\langle1|,\quad
	J_5= \sqrt{\gamma_3}|\chi_2\rangle\!\langle2| ,
\end{align}
where $|\chi_1\rangle = \cos\theta|0\rangle+\sin\theta|2\rangle$ and $|\chi_2\rangle=-\sin\theta|0\rangle+\cos\theta|2\rangle$.
Then, the SJEDs are $S_1=\{ 1,2,3\}$ and $S_2 = \{4,5\}$, they are both of reset type.  Their composite actions are
\begin{equation}
	\mathcal{A}_1 (\psi)=\Tr(\Gamma\psi)\, |\chi_1\rangle\!\langle\chi_1|, \quad \mathcal{A}_2 (\psi) = \Tr(\Gamma\psi)\,|\chi_2\rangle\!\langle\chi_2|,
\end{equation}
with $\Gamma =(\gamma_1+\gamma_2)|1\rangle\!\langle1|+\gamma_3|2\rangle\!\langle2|$.

For a second representation, we consider the same Hamiltonian $H$, and jump operators 
\begin{align}
	\tilde  J_1 &= \sqrt{\tilde \gamma_1}|0\rangle\!\langle1|, \quad 
	\tilde  J_2 = \sqrt{\tilde \gamma_2}|0\rangle\!\langle1| , \quad
	\tilde J_3 = \sqrt{\gamma_3}|0\rangle\!\langle2|,\nonumber\\
	\tilde J_4 &= \sqrt{\tilde \gamma_1+\tilde \gamma_2}|2\rangle\!\langle1|, \quad
	\tilde J_5 = \sqrt{\gamma_3}|2\rangle\!\langle2|,
\end{align}
where $\tilde \gamma_1+\tilde \gamma_2=\gamma_1+\gamma_2$ to ensure the same QME.  
The SJEDs are $\tilde S_1=\{ 1,2,3\}$ and $\tilde S_2 = \{4,5\}$, again both of reset type. 
Their composite actions are
\begin{equation}
	\mathcal{A}_1 (\psi)=\Tr(\Gamma\psi)\, |0\rangle\!\langle0|, \quad \mathcal{A}_2 (\psi)= \Tr(\Gamma\psi)\,|2\rangle\!\langle2|.
\end{equation}

The two representations $H,J_1,\dots,J_5$ and  $H,\tilde J_1,\dots,\tilde J_5$ do not generally satisfy Theorem~1 as their SJEDs differ in their reset states,  but there are special cases. For example:
\begin{align}
&\theta= 0^\circ :  &\hspace{-35mm} \tilde{\cal A}_1 = {\cal A}_1 , \qquad  \tilde{\cal A}_2 = {\cal A}_2,
	\nonumber\\
&\theta= 90^\circ :  &\hspace{-35mm} \tilde{\cal A}_1 = {\cal A}_2 , \qquad  \tilde{\cal A}_2 = {\cal A}_1.
\end{align}
Theorems~1 and~3 are valid in both these cases,  which means that the two representations give rise to the same ensemble of quantum trajectories.  Their partially labelled quantum trajectories are identical for $\theta= 0^\circ$, and equivalent for $\theta=90^\circ$ (as the SJED labels are swapped). 

For these two representations, 
Theorem~2 only applies only if $\theta=0$ and $\tilde\gamma_1=\gamma_1$ (which also implies $\tilde\gamma_2=\gamma_2$). Then, the  permutation in Theorem~2 is trivial as $\tilde J_k=J_k$ for $k=1,...,d$.  If additionally $\gamma_1=\gamma_2$ holds (which also implies $\tilde\gamma_1=\tilde\gamma_2$), then $J_1=J_2$ and $\tilde{J}_1=\tilde{J}_2$.  In this case the permutation in Eq.~\eqref{eq:WF_cond} is no longer unique: it can be chosen either as trivial or to swap $1$ and $2$ types.
\medskip

\section{Pure state with distinct destinations for all SJEDs}\label{app:C}
Here, we describe how to find $\psi_0$ which is at the center of $\cset$  in Eq.~\eqref{eq:C} that is used in the proof of Theorem~1.
This discussion refers back to the properties (i)-(iii) of set $\cset$, given at the beginning of Sec.~\ref{sec:different-dest}.

Start from an arbitrary candidate state $\psi=|\psi\rangle\!\langle\psi|$.  (In the following, we sometimes denote states via matrices $\psi$ and sometimes via the corresponding vectors $|\psi\rangle$).  Suppose that $\psi$ violates the conditions (i,ii) above, because of two ``degeneracies'', that $\mathcal{D}_k(\psi)=0$ and $\mathcal{D}_{k'}(\psi)=\mathcal{D}_{k''}(\psi)\neq0$ for some $k,k',k''$ (all in different SJEDs). 
Then choose some $|\varphi\rangle$ such that $J_k|\varphi\rangle\neq0$ and write $|\psi'\rangle=(|\psi\rangle+a|\varphi\rangle)/z$ with $a\in\mathbb{R}$, where $z$ is a normalisation constant.  Then $\mathcal{D}_k(\psi')\neq0$, so condition (i) is now satisfied.  Moreover, taking sufficiently small $a>0$  ensures that this replacement does not generate any new degeneracies [for example, it avoids the situation that some $\mathcal{D}_{k'''}(\psi)\neq 0$ but  $\mathcal{D}_{k'''}(\psi')= 0$].  

The degeneracy $\mathcal{D}_{k'}(\psi')=\mathcal{D}_{k''}(\psi')\neq0$ may still remain, in which case we write $|\psi''\rangle=(|\psi'\rangle+a'|\varphi'\rangle)/z'$ with a new constant $a'$ and some pure state $\varphi'$ such that $\mathcal{D}_{k'}(\varphi')\neq \mathcal{D}_{k''}(\varphi')$ [such a state always exists because otherwise $k',k''$ would be in the same SJED].  Again, one may take $a'$ small enough that no new degeneracies are created [including that we still have $\mathcal{D}_k(\psi'')\neq0$].  This two-step process eliminates both the degeneracies of $\psi$ so one may take $\psi_0=\psi''$, {which now satisfies conditions (i) and (ii). The same construction can be performed when considering degeneracies in the $\tilde{\mathcal{D}}_j(\psi)$ which violate conditions (i) and/or (iii)}.   

If the initial candidate had more than two degeneracies, one would repeat the same method some (finite) number of times, in order to remove them all.  In this way, a suitable $\psi_0$ can always be constructed.

\bibliographystyle{quantum}
\bibliography{quantum_refs}

\end{document}